\documentclass[a4paper,11pt]{article}
\usepackage{graphicx} 
\usepackage[utf8]{inputenc}
\usepackage[T1]{fontenc}
\usepackage{comment}
\usepackage[T1]{fontenc}
\usepackage{subfigure}
\usepackage{float}
\usepackage{natbib}
\usepackage{epstopdf}
\usepackage{xcolor}
\usepackage{a4wide}
\usepackage{indentfirst}
\usepackage{authblk}
\usepackage{booktabs}

\usepackage{tikz}
\usetikzlibrary{patterns}
\usetikzlibrary{arrows.meta,positioning}
\usepackage{standalone}

\usepackage{amsmath,amssymb}
\usepackage{amsthm}
\usepackage{mathtools}

\usepackage[margin=1.2in]{geometry}
\usepackage{graphicx}
\usepackage{subcaption}
\usepackage{listings}
\usepackage[utf8]{inputenc}
\usepackage{appendix}
\usepackage{hyperref}
\definecolor{turq}{RGB}{0,150,170}
\hypersetup{
  colorlinks=true,
  citecolor=turq,
  linkcolor=black,
  urlcolor=black
}
\numberwithin{equation}{section}
\newtheorem{teo}{Theorem}[section]
\newtheorem*{teo*}{Theorem}
\newtheorem*{prop*}{Proposition}
\newtheorem*{corol*}{Corollary}
\newtheorem{prop}[teo]{Proposition}

\newtheorem{lema}[teo]{Lemma}
\newtheorem{defi}[teo]{Definition}

\theoremstyle{definition}

\newcommand{\Cov}{\operatorname{Cov}}
\renewcommand{\d}{\,\mathrm{d}}

\newcommand{\D}{\mathbb{D}}
\newcommand{\E}{\mathbb{E}}
\newcommand{\F}{\mathcal{F}}

\renewcommand{\P}{\mathbb{P}}

\newcommand{\R}{\mathbb{R}}
\newcommand{\Var}{\operatorname{Var}}
\newcommand{\1}{\mathbf{1}}
\newcommand{\phy}{\varphi}
\newcommand{\eps}{\varepsilon}

\title{ Gatheral's Conjecture: General sufficient conditions for refutation}
\author[1]{Elisa Alòs}
\author[2]{Òscar Burés}
\author[3]{Vladimir Lucic}

\affil[1]{Universitat Pompeu Fabra and Barcelona School of Economics, Department of Economics and Business, Barcelona, Spain, \vspace*{3pt}}
\affil[2]{ Universitat de Barcelona, Departament de Matemàtica Econòmica, Financera i Actuarial. \authorcr Barcelona, Spain, \vspace*{3pt}}
\affil[3]{Imperial College London, Department of Mathematics, South Keningston Campus}
\date{\today}
\begin{document}

\maketitle

\begin{abstract}
    In this paper we give general sufficient conditions for a stochastic volatility model to be a counterexample for Gatheral's conjecture. The conditions, achieved by asymptotic properties of integrated variance derivatives and Wiener chaos expansions, show that a general class of rough volatility models are generally counterexamples to Gatheral's conjecture. For stochastic volatility models with Markovian variance, the conjecture fails under a relationship on the coefficients of the SDE defining the variance process. Particular examples are provided.
\end{abstract}
\section{Introduction}
Gatheral’s conjecture, stated in \cite{gatheral2005valuation}, reads as follows: “Given the prices of European options of all strikes and expirations, of all possible underlying diffusions consistent with these option prices, the lowest possible value of a volatility option is achieved by assuming local volatility dynamics”. Earlier practitioner’s observations, such as \cite{brockhaus2000volatility}, reported higher fair volatility swap strikes in a calibrated local volatility model rather than a Heston model. Since the volatility swap fair strike is an example of a concave payoff, the observation was in line with Gatheral’s conjecture.

Soon after the formalization of the conjecture, \cite{beiglbock2011minimum} gave two counterexamples that settled the conjecture as false. Nevertheless, the examples provided are not the ones of stochastic volatility models commonly used in quantitative finance literature, so the natural question that remained open from their work is whether well-known stochastic volatility models do refute Gatheral’s conjecture.

Recently, \cite{lucic2026gatheral} addresses this question for the Heston model with perfectly negative spot-variance correlation. In particular, not only the author shows that such exampe does refute Gatheral’s conjecture, but the prices of European call options on the integrated variance under the Heston  model with perfect negative correlation remain uniformly lower than the same prices with the local integrated variance as underlying asset. The argument given in the paper, however, relies heavily on the Heston model with perfect negative correlation structure, so the arguments used in the paper are difficult to replicate to other stochastic volatility models. 

Motivated by the result in \cite{lucic2026gatheral}, in this paper we give general conditions for which a stochastic volatility model does refute Gatheral's conjecture. In particular, for a general class of stochastic volatility models, we give conditions on the model’s parameters in order to refute Gatheral’s conjecture if the time to maturity is small enough. This paper constitutes therefore the natural continuation of \cite{lucic2026gatheral} since, in particular, it answers some of the open questions stated in the paper.
 To provide a general structure for a stochastic volatility model, we let the variance process satisfy a Volterra SDE with Riemann-Liouville kernel and Hurst index $H\leq \frac{1}{2}$. In the paper, we show that if $H <1/2$ and the correlation is, in absolute value, greater than an explicit threshold depending on $H$ then Gatheral’s conjecture is refuted. For $H = \frac{1}{2}$ the variance process satisfies an ordinary SDE. In this case, we give a condition on the coefficients  in order to refute Gatheral’s conjecture provided $|\rho| = 1$ and the expiry time is small enough. Our result covers a broad class of examples. In particular, a direct application of our result shows that for a big enough correlation (in absolute value), a general class of rough volatility models, the Heston model under certain parameter constraints and the SABR model refute Gatheral’s conjecture.

The paper is structured as follows: Section \ref{sec: preliminaries} is devoted to the formulation of the problem together with our main result, Theorem \ref{th: main}. In Section \ref{sec: Malliavin}, we give the necessary Malliavin calculus tools in order to prove Theorem \ref{th: main}. Sections \ref{sec: rough} and \ref{sec: Markovian case} are devoted to proof both statements in Theorem \ref{th: main}. Finally, in Section \ref{sec: loc and examples} we show that the hypotheses of Theorem \ref{th: main} can be relaxed to cover a broader class of examples, and we exemplify the result with the rough Heston model, the classical Heston model and the SABR model.

\section{Preliminaries} \label{sec: preliminaries}
Throughout the paper, we will assume for simplicity null interest rates, $r = 0$. Let $S = \{S_t; t \in [0,T]\}$ be the asset price process, defined as
\begin{equation} \label{eq: original model}
\frac{dS_t}{S_t} = \sqrt{v_t} (\rho dW_t + \sqrt{1-\rho^2}dB_t),
\end{equation}
where $W$ and $B$ are independent Brownian motions defined in a complete probability space $(\Omega, \F, \P)$. Here, $v_t \geq 0$ denotes the variance process, which we assume it measurable with respect to the natural filtration generated by $W$ and $\rho \in [-1,1]$. Let$\lambda(t,X_t) = \E[v_t|X_t]$ with $X_t = \log S_t$ denote the projected local variance and let $\overline{S}$ defined by
\begin{equation} \label{eq: local variance model}
\frac{d\overline{S}_t}{\overline{S}_t}= \sqrt{\lambda(t,\overline{X}_t)}d\hat{W}_t, \quad \overline{X}_t = \log \overline{S}_t,
\end{equation}
denote the local variance model associated to the stochastic volatility model given by \eqref{eq: original model}. Given \cite[Corollary 3.7]{BrunickShreve2013}, if $v_t$ satisfies $\E[\int_0^T v_t dt] < \infty$ then there exists a weak solution $\overline{S}$ to Equation \eqref{eq: local variance model} such that the one-dimensional marginals of $S$ and $\overline{S}$ coincide.

We define the integrated variance of both models \eqref{eq: original model} and \eqref{eq: local variance model} respectively as
\begin{equation} \label{eq: integrated variances definition}
I_T = \int_0^T v_t dt, \quad I^{LV}_T = \int_0^T \lambda(t,\overline{X}_t) dt.
\end{equation}
Gatheral's conjecture, stated from the mathematical point of view, says that for every convex payoff $\phy$ and every expiry time $T$, one has
\[
\E[\phy(I_T^{LV})] \leq \E[\phy(I_T)].
\]
In order to prove that the conjecture is false one must find a convex payoff $\phy$ such that for an expiry time $T$,
\begin{equation} \label{eq: Gatheral condition}
\E[\phy(I_T)] < \E[\phy(I_T^{LV})].
\end{equation}
We refer to the property \eqref{eq: Gatheral condition} as Gatheral's condition. In this note we describe a family of models such that Gatheral's condition is not satisfied for $T$ close to zero, and therefore they constitute counterexamples to Gatheral's conjecture. To do so, fix a general family of variances of the form
\begin{equation} \label{eq: variance model}
v_t = v_0 + \int_0^t K(t-s) \alpha(v_s) ds + \int_0^t K(t-s)\beta(v_s) dW_s, \quad K(t-s) = \frac{(t-s)^{H-\frac{1}{2}}}{\Gamma(H + \frac{1}{2})}.
\end{equation}
with $H \in (0, \frac{1}{2}]$. In order to give conditions for which Gatheral's condition \ref{eq: Gatheral condition} under a variance model the form \eqref{eq: variance model}, very relaxed hypotheses on the coefficients $\alpha$ and $\beta$ are needed. However, the proof of the result is easier if we conduct a localization argument. This means that, for the moment, we will assume very strong hypotheses on $\alpha$ and $\beta$ and then, in Section \ref{sec: loc and examples} we will show how we can relax the hypotheses while the main result holds true. 

We assume the following hypotheses on $\alpha$, $\beta$ and the local variance.
\begin{itemize}
    \item [(H1)] There exists $0 < v_{min} < v_{max}$ such that
    \[
    v_{min} \leq v_t \leq v_{max}
    \]
    for all $t \in [0,T]$.
    \item [(H2)] $\alpha, \beta \in \mathcal{C}^{\infty}([0,\infty))$, with $\beta(v_0) \neq 0$.
    \item [(H3)]
    The function $\lambda(t,x)$ satisfies
    \[
    \partial_x^n \lambda(t, \overline{X}_t) \leq Ct^{n(H-\frac{1}{2})}.
    \]
\end{itemize}
Our main result gives conditions on $\rho$, $H$ and the coefficients $\alpha$ and $\beta$ of the variance equation \eqref{eq: variance model} to obtain violations of Gatheral's property.
\begin{teo} \label{th: main}
    Let $I_T$ and $I^{LV}_T$ be defined as in \eqref{eq: integrated variances definition}. Then, under Hypotheses (H1)--(H3) we have:
    \begin{itemize}
        \item[(i)] If $H \in (0,\frac{1}{2})$, then
        \[
        \E\left[\left(I_T - K^* \right)_+\right] <   \E\left[\left(I_T^{LV} - K^* \right)_+\right]
        \]
        whenever
        \[
            \rho^2 > \frac{2H+3}{4}.
        \]
        Here $K^* = \E[I_T]$ denotes the at-the-money strike.
        \item[(ii)] If $H = \frac{1}{2}$ and $|\rho| = 1$, then
        \[
        \E\left[I_T^2 \right] < \E\left[(I_T^{LV})^2 \right]
        \]
        whenever
        \[
          K_0(\alpha,\beta, \rho) = \alpha'_0 \beta_0^2+ \frac{\alpha_0\beta_0^2}{2v_0} - \frac{\beta_0^4}{8v_0^2} + \frac{\rho \beta_0^3}{2\sigma_0}-\alpha_0 \beta_0 \beta_0' - \frac{\beta_0^3\beta_0^{''}}{2} < 0.
        \]
    \end{itemize}
    In particular, the two cases stated before are counterexamples to Gatheral's conjecture.
\end{teo}
The proof of the two previous items are done separately, since two different strategies are applied. For the rough case, it will be enough to resort to results on derivatives on realized variance. For the Markovian case, we will have to resort to Wiener chaos expansions, as it is a suitable tool in order conduct an analysis under the $|\rho| = 1$ condition due to the fact that both $X_t$ and $v_t$ are in that case defined on the same Wiener space. In the next section, we provide the Malliavin calculus tools needed for proving Theorem \ref{th: main}.

\section{Malliavin Calculus} \label{sec: Malliavin}
This section is devoted to provide a primer on the Malliavin calculus tools we have used in the paper. For a complete reference, we refer the reader to \cite{nualart2006}. Let $W = \{W_t; t \in [0,T]\}$ be a standard Brownian motion defined in a complete probability space $(\Omega, \F, \mathbb{P})$. For a function $h \in L^2([0,T])$ we define $W(h)$ as the Wiener integral of $h$ with respect to $W$, i.e.
\[
W(h) := \int_0^T h(t) \d W_t.
\]
We define $\mathcal{S}$ as the class of random variables of $F$ the form
\[
F = f(W(h_1), \dots, W(h_n)); \quad h_i \in L^2([0,T]), \quad f\in \mathcal{C}^{\infty}_b(\R^n),
\]
where $\mathcal{C}_b^{\infty}(\R^n)$ denotes the space of continuous and bounded functions which are infinitely many times differentiable and $f$, together with its partial derivatives, are continuous and bounded. We say that $\mathcal{S}$ is the space of \textit{simple functionals}. 
\begin{defi}
    Let $F \in \mathcal{S}$. We define the Malliavin derivative of $F$ evaluated at a point $t \in [0,T]$ as the stochastic process defined by
    \[
    D_t F = \sum_{j=1}^n \partial_j f(W(h_1), \dots, W(h_n)) h_j(t), \quad \partial_j f := \frac{\partial f}{\partial x_j}.
    \]
\end{defi}
The operator $D$ is closable from $L^p(\Omega)$ to $L^p(\Omega \times [0,T])$ for all $p \geq 1$ (see \cite{nualart2006}). We define $\D^{1,p}_{W}$ as the closure of $\mathcal{S}$ with respect to the norm
\[
||F||_{1,p} = \left[ \E(|F|^p) + \E\left(||DF||_{L^2([0,T])}^p\right)\right]^{1/p}.
\]
In the same way we have defined $D$, we can define the iterated derivatives $D^n$. We denote $\D^{n,p}_{W}$ the space of random variables that are $n$ times Malliavin differentiable and $F$, together with its Malliavin derivatives are $p$-integrable.
\subsection{Wiener chaos expansions}
Any random variable $F$ that is $\mathcal{F}_T$-measurable can be expanded in a series of multiple stochastic integrals
\[
F = \E[F] + \sum_{n=1}^{\infty} I_n(f_n)
\]
where $f_n \in L^2_{sym}([0,T]^n)$ and $I_n(f_n)$ denotes the $n$-th iterated Wiener integral of $f$. We refer the reader to \cite{nualart2006} for a further treatment of this topic. In terms of the chaos expansions, the variance of $F$ can be computed as
\[
\Var(F) = \sum_{n=1}^{\infty} n! \int_{[0,T]^n}|f_n(z)|^2 dz = \sum_{n=1}^{\infty} n! ||f_n||^2_{L^2([0,T]^n)}.
\]
If $F \in \bigcap_{k \geq 1} \D^{k,2}$ we have
\begin{equation} \label{eq: stroock}
\Var(F) =\sum_{n=1}^{\infty} \frac{1}{n!}||\E[D_{\cdot}^n F]||^2_{L^2([0,T]^n)}.
\end{equation}
Moreover, the following Poincaré-type inequality (see, for instance, \cite{chen2021spatial} Equation (1.15))  holds:
\begin{equation} \label{eq: Poincare}
    \Var(F) \leq \int_0^T\E[|D_uF|^2] du.
\end{equation}
 
\section{The rough case} \label{sec: rough}
In this section we will provide the proof of Theorem \ref{th: main}-(i), which means that we focus on $v$ satisfying Equation \eqref{eq: variance model} with $H < \frac{1}{2}$. The strategy for the rough regime is rewriting Gatheral's condition as a problem of comparing implied volatilities. Towards this end, we define $IV(T,K)$ the implied volatility of the European call options with the integrated variance as underlying asset, time to maturity $T$ and strike $K$. For the local variance model, we write $IV^{LV}(T,K)$. The problem of determining under which conditions we have
\[
\E[(I_T - K)_+]<\E[(I_T^{LV}-K)_+]
\]
is equivalent of finding conditions in order to have
\[
\frac{IV(T,K)}{IV^{LV}(T,K)} < 1.
\]
In \cite{alosGLmuguruza}, the authors provide a framework for studying ATM implied volatilities of integrated variance derivatives. Indeed, write $M_t = \E[I_T |\mathcal{F}^W _t]$. Then, the martingale representation theorem shows that we can write
\[
M_t = M_0 + \int_0^t m(T,s) dW_s.
\]
Let $\phi_t := m(T,t)M_t^{-1}$. Then, the authors in \cite{alosGLmuguruza} show that if the following limit holds
\[
\lim_{T \to 0}\frac{1}{T^H} \E\sqrt{\int_0^T \phi_T(t)^2 dt} = \ell,
\]
then
\[
\lim_{T \to 0} T^{1/2 - H} I(T,\E[I_T]) = \ell.
\]
The strategy in order to prove Theorem \ref{th: main}-(i) is computing $\ell$ for the stochastic volatility model and the local variance model respectively, and analyzing the quotient to see under which conditions it is strictly smaller than one. In order to compute such limits, we will use the same argument for both models. First, we rewrite the limit as
\[
\ell = \lim_{T \to 0}\frac{1}{T^H} \E[||\phi_T(\cdot)||_{L^2(0,T)}].
\]
Let $\overline{\phi}_T(t)$ denote a deterministic function. Then the Cauchy-Schwarz inequality yields
\[
\frac{1}{T^H} \left(\E[||\phi_T(\cdot)||_{L^2(0,T)}] -||\overline{\phi}_T(\cdot) ||_{L^2(0,T)}\right) \leq T^{\frac{1}{2}-H}\sup_{t \in [0,T]}||\phi_T(t) - \overline{\phi}_T(t)||_{L^2(\Omega)}.
\]
If we are able to find a deterministic function $\overline{\phi}_T(t)$ such that
\[
\sup_{t \in [0,T]}||\phi_T(t) - \overline{\phi}_T(t)||_{L^2(\Omega)} = o(T^{H-\frac{1}{2}})
\]
and the limit $T^{-H} ||\overline{\phi}_T(\cdot)||_{L^2(\Omega)} \to \ell$ as $T \to 0$ we can conclude that $\ell = \lim_{T \to 0} T^{-H} \E[||\phi_T(\cdot)||_{L^2(0,T)}]$. In order to find candidates for $\overline{\phi}_T(t)$, we use that in short-times we must heuristically have $M_t \approx v_0 T$ in order to find a candidate of the form
\[
\overline{\phi}_T(t) = \frac{\overline{m}(T,t)}{v_0T},
\]
In fact, the following result shows that the accuracy of the approximation does only depend on $\overline{m}(T,t)$.
\begin{lema} \label{lemma: acc on m}
    Under Hypotheses (H1)--(H3) we have that
    \[
    ||v_t - v_0||_{L^2(\Omega)} \leq Ct^{H}, \quad ||\lambda(t,\overline{X}_t) - v_0||_{L^2(\Omega)} \leq Ct^{H},
    \]
    and therefore,
    \[
    ||\phi_T(t) - \frac{\overline{m}(T,t)}{v_0 T}||_{L^2(\Omega)} \leq C\left( \frac{1}{T}||m(T,t) - \overline{m}(T,t)||_{L^2(\Omega)} + T^{H-1}|\overline{m}(T,t)| \right).
    \]
\end{lema}
\begin{proof}
    The assertion $||v_t-v_0||_{L^2(\Omega)} \leq Ct^{H}$ follows by the hypotheses on $\alpha$ and $\beta$, together with Burkholder-Davis-Gundy inequality. For the local variance, we use the mean value theorem to say
    \[
    \lambda(t,\overline{X}_t) - v_0 = \lambda(t,\overline{X}_t) - \lambda(t,X_0) + \lambda(t,X_0) - v_0.
    \]
    On the one hand, the mean value theorem yields
    \[
    ||\lambda(t,\overline{X}_t) - \lambda(t,X_0)||_{L^2(\Omega)} \leq Ct^{H-\frac{1}{2}}||\overline{X}_t - X_0||_{L^2(\Omega)} \leq Ct^{H}.
    \]
    On the other hand,
    \[
    ||\lambda(t,X_0) - v_0||_{L^2(\Omega)} \leq ||\lambda(t,\overline{X}_t) - \lambda(t,X_0)||_{L^2(\Omega)} + ||v_t- v_0||_{L^2(\Omega)} \leq Ct^{H},
    \]
    where we have used that $\lambda(t,X_0) - v_0 = \E[\lambda(t,X_0) - \lambda(t,X_t)] + \E[v_t - v_0]$. 

    Now, Minkowski's inequality shows that
    \[
    ||M_t -v_0T||_{L^2(\Omega)} \leq \int_0^T ||v_s-v_0||_{L^2(\Omega)} ds \leq C T^{H+1}.
    \]
    Hence,
    \[
    ||\phi_T(t) - \frac{\overline{m}(T,t)}{v_0T}||_{L^2(\Omega)} \leq ||\frac{m(T,t) - \overline{m}(T,t)}{M_t}||_{L^2(\Omega)} +  |\overline{m}(T,t)| \cdot ||\frac{M_t-v_0T}{M_tv_0T} ||_{L^2(\Omega)}.
    \]
    Using the previous bounds, and the fact that $|M_t| \geq C T$ we have
    \[
        ||\phi_T(t) - \frac{\overline{m}(T,t)}{v_0 T}||_{L^2(\Omega)} \leq C\left( \frac{1}{T}||m(T,t) - \overline{m}(T,t)||_{L^2(\Omega)} + T^{H-1}|m(T,t)| \right),
    \]
    as desired.
\end{proof}
The rest of the section is devoted to finding $\overline{m}$ for both the stochastic volatility and local variance model, and testing the goodness of the candidate $\overline{\phi}_T(t)$.
\subsection{The stochastic volatility limit}
For the stochastic volatility model we consider
\[
\overline{m}(T,t) =  \frac{\beta_0}{\Gamma(H+3/2)}(T-t)^{H+\frac{1}{2}}.
\]
The goodness of the candidate resides in the following result.
\begin{prop} \label{prop: estimation m SV}
    Under Hypotheses (H1)--(H3) we have
    \[
    ||m(T,t) - \overline{m}(T,t)||_{L^2(\Omega)} \leq CT^{2H + \frac{1}{2}}.
    \]
\end{prop}
\begin{proof}
    Fist notice that
    \[
    \overline{m}(T,t) = \beta_0 \int_t^T K(s-t) ds.
    \]
    Then, the Clark-Ocone formula allows us to write
    \[
    m(T,t) - \overline{m}(T,t) = \int_t^T \left(\E_t[D_tv_s] - \beta_0K(s-t) \right) ds.
    \]
    Hence, by Minkowski's inequality it is enough to bound
    \[
    ||\E_t[D_tv_s] - \beta_0K(s-t) ||_{L^2(\Omega)}.
    \]
    The Malliavin derivative of $v$ satisfies the following equation
    \[
    D_tv_s = \beta(v_t) K(s-t) + \int_t^s K(s-u)\alpha'(v_u)D_tv_u du + \int_t^s K(s-u)\beta'(v_u) D_tv_u dW_u.
    \]
    Hence,
    \begin{align*}
        ||\E_t[D_tv_s] - \beta_0K(s-t)||_{L^2(\Omega)} \leq &CK(s-t)||v_t - v_0||_{L^2(\Omega)} \\
        + & C\int_t^s K(s-u) ||D_tv_u||_{L^2(\Omega)} du
    \end{align*}
    Using the same arguments as in \cite{bures2026skew} we have $||D_tv_s||_{L^p(\Omega)} \leq C_p K(s-t)$. Using also that $||v_t-v_0||_{L^2(\Omega)} \leq Ct^{H}$ and integrating,
    \[
     ||\E_t[D_tv_s] - \beta_0K(s-t)||_{L^2(\Omega)} \leq  C\left(t^{H}K(s-t) + \frac{(s-t)^{2H}}{\Gamma(2H+1)}\right).
    \]
    Finally,
    \[
    || m(T,t) - \overline{m}(T,t) ||_{L^2(\Omega)} \leq \int_t^T   ||\E_t[D_tv_s] - \beta_0K(s-t) ||_{L^2(\Omega)} ds \leq CT^{2H+\frac{1}{2}}.
    \]
\end{proof}
Using the fact that $|\overline{m}(T,t)| = O(T^{H+\frac{1}{2}})$ and the estimate provided in Proposition \ref{prop: estimation m SV}, we have
\[
\ell = \lim_{T \to 0} \frac{1}{T^H}||\overline{\phi}_T(\cdot)||_{L^2(0,T)} = \frac{\beta_0}{v_0 \Gamma(H+3/2)\sqrt{2H+2}}.
\]

\subsection{The local variance limit}
In order to redo the computations for the local variance model we first need to make sure that $\overline{X}$ exists as a strong solution to its defining SDE. The existence and uniqueness of $\overline{X}$, together with its Malliavin differentiability is discussed in Section \ref{sec: local variance model}. Let
$\sigma(t,\overline{X}_t) = \sqrt{\lambda(t,\overline{X}_t)}$. The Malliavin derivative of $\overline{X}_t$ can be written as
\[
D_t \overline{X}_s = \sigma(t,\overline{X}_t) Y_{t,s}, \quad Y_{t,s} = \exp\left\{ \int_t^s \partial_x\sigma(u,\overline{X}_u) d\hat{W}_u - \int_t^s (\sigma \partial_x \sigma + \frac{1}{2}(\partial_x\sigma)^2)(u, \overline{X}_u) du\right\}.
\]
For the local variance model, we choose the candidate
\[
\overline{m}^{LV}(T,t) = \frac{\rho \beta_0}{\Gamma(H+3/2)(H+1/2)}\left(T^{H+\frac{1}{2}}-t^{H+\frac{1}{2}} \right).
\]
We again prove the goodness of the candidate via the following result.
\begin{prop} \label{prop: estimation m LV}
    Under Hypotheses (H1)--(H3) we have
    \[
    ||m(T,t) - \overline{m}^{LV}(T,t)||_{L^2(\Omega)} \leq C (T^H + \eps(T))T^{H+\frac{1}{2}},
    \]
    where $\eps(T) \to 0$ as $T \to 0$.
\end{prop}
\begin{proof}
    First, notice that
    \[
    \overline{m}^{LV}(T,t) =  \frac{\rho \beta_0}{\Gamma(H+3/2)} \int_t^T s^{H-\frac{1}{2}}ds. 
    \]
    On the other hand, using the definition $\sigma = \sqrt{\lambda}$, and letting
    \[
    c_{LV}:=  \frac{\rho \beta_0}{\Gamma(H+3/2)},
    \]
    we have
    \begin{align*}
    ||m^{LV}(T,t) - \overline{m}^{LV}(T,t) ||_{L^2(\Omega)} \leq &\int_t^T ||\partial_x\lambda(s,\overline{X}_s) D_t\overline{X}_s - c_{LV}s^{H-\frac{1}{2}}||_{L^2(\Omega)}ds \\
    = & \int_t^T s^{H-\frac{1}{2}} ||s^{\frac{1}{2}-H}\partial_x \lambda(s,\overline{X}_s) D_t\overline{X}_s  -c_{LV}||_{L^2(\Omega)}ds.
    \end{align*}
    Due to the $\frac{1}{H+3/2}$-rule in \cite{AlosGLPravosud} and the short-time expansion of the ATM-IV skew as in \cite{alos2007short} in the particular case of absence of jumps we have
    \[
    \lim_{s \to 0}s^{\frac{1}{2}-H}\partial_x\lambda(s,\overline{X}_s) = \lim_{s \to 0} 2\sigma(s,\overline{X}_s)s^{\frac{1}{2}-H} \partial_x \sigma(s,\overline{X}_s) = \frac{\rho\beta_0}{\sigma_0 \Gamma(H+3/2)} = \frac{c_{LV}}{\sigma_0}.
    \]
    Now, consider $D_t\overline{X}_s = \sigma(t,\overline{X}_t) Y_{t,s}$. Write also $Z_{t,s} = \log Y_{t,s}$. Then, using that $|e^x-1| \leq |x|(1+e^x)$ we have
    \[
    ||D_t\overline{X}_s - \sigma_0||_{L^2(\Omega)} \leq ||\sigma(t,\overline{X}_t) - \sigma_0||_{L^4(\Omega)}||Y_{t,s}||_{L^4(\Omega)} + \sigma_0||Z_{t,s}||_{L^4(\Omega)}(1+||Y_{t,s}||_{L^4(\Omega)}).
    \]
    On the one hand, the fact that $||\lambda(t,\overline{X}_t) - v_0||_{L^4(\Omega)} \leq Ct^{H}$ implies that
    \[
    ||\sigma(t,\overline{X}_t) - \sigma_0||_{L^4(\Omega)} \leq Ct^{H}.
    \]
    A classical argument with Burkholder-Davis-Gundy inequality shows that $||Z_{t,s}||_{L^4(\Omega)} \leq Cs^{H}$ and that the moments of $Y_{t,s}$ are finite. Then,
    \[
      ||D_t\overline{X}_s - \sigma_0||_{L^2(\Omega)} \leq Cs^H.
    \]
  We split the estimation into two contributions, namely
  \begin{align*}
  &||s^{\frac{1}{2}-H}\partial_x \lambda(s,\overline{X}_s) D_t\overline{X}_s  -c_{LV}||_{L^2(\Omega)}\\
  \leq &||s^{\frac{1}{2}-H}\partial_x \lambda(s,\overline{X}_s)(D_t\overline{X}_s- \sigma_0)||_{L^2(\Omega)} + \sigma_0 ||s^{\frac{1}{2}-H}\partial_x \lambda(s,\overline{X}_s)-\frac{c_{LV}}{\sigma_0}||_{L^2(\Omega)} \\
  \leq & C \left(s^{H} +  \eps(s) \right),
  \end{align*}
  so
  \[
  |m^{LV}(T,t) - \overline{m}^{LV}(T,t) ||_{L^2(\Omega)} \leq  C\int_t^T (s^{2H-\frac{1}{2}} + s^{H-\frac{1}{2}}\eps(s)) ds 
  \]
  which is of order $O(T^{2H+\frac{1}{2}}) + o(T^{H+\frac{1}{2}})$, concluding the result.
\end{proof}
Proposition \ref{prop: estimation m LV} allows us to write
\[
\ell^{LV} = \lim_{T \to 0} \frac{1}{T^H}||\overline{\phi}_T^{LV}(\cdot))||_{L^2(0,T)} = \frac{|\rho| \beta_0}{v_0\Gamma(H+3/2) \sqrt{(H+1)(H+3/2)}}.
\]
We are now ready to prove Theorem \ref{th: main}- (i)

\begin{proof}{(of Theorem \ref{th: main}-(i))}
    We recall that in order to show that 
    \[
    \E[(I_T - K^*)] < \E[(I_T^{LV} - K^*)_+],
    \]
    where $K^* = \E[I_T]$ we can show that
    \[
    \frac{IV(T,K^*)}{IV^{LV}(T,K^*)}< 1.
    \]
    Taking limits as $T \to 0$, we have
    \[
    \lim_{T \to 0} \frac{IV(T,K^*)}{IV^{LV}(T,K^*)} = \frac{\ell}{\ell^{LV}} = \frac{\sqrt{(H+1)(H+3/2)}}{|\rho|\sqrt{2(H+1)}} = \frac{\sqrt{2H+3}}{2|\rho|}.
    \]
    Hence, if
    \[
    \rho^2 > \frac{2H+3}{4}
    \]
    there exists $T_0$ small enough such that the Gatheral property fails for the quadratic payoff $\phy(x) = (x-K^*)_+$ for every maturity $T \in (0,T_0]$.
\end{proof}
\section{The Markovian case} \label{sec: Markovian case}
Notice that when $H = \frac{1}{2}$, the condition of Theorem \ref{th: main}-(i) becomes $\rho^2 > 1$ which can not hold. In order to give conditions on the $H = \frac{1}{2}$ case, we will consider $|\rho| = 1$ and we will work with the quadratic payoff with null strike $\phy(x) = x^2$. The reason behind the choice of this payoff is that, in the case where $|\rho| = 1$ both $v$ and $X$ admit Wiener chaos expansions in the same chaos spaces. Since the variance of a random variable is related to the sum of the norms of the projections to the Wiener chaos' spaces, one can consider the problem of the difference of the variances as a problem of the difference of two series derived from the Wiener chaos expansions of $I_T$ and $I_T^{LV}$. 

\subsection{Eliminating the dependence on $\overline{X}$}
It is not convenient for the Markovian case to study directly the interaction between $v$ and $\lambda(t,\overline{X}_t)$ because we do not know enough about $\lambda(t,\overline{X}_t)$ in order to provide a short-time chaos expansion. Nevertheless, we can provide a short-time chaos expansion for $X_t$, so the first objective in order to reach the result is showing that we can reformulate the problem so that $\overline{X}$ does not participate in any step of the proof. To do so, we follow closely the arguments in \cite{lucic2026gatheral}.

Let $\phi$ denote a regularization of the payoff $z \mapsto z^2$ constructed in the following way: Let $M$ such that $\max(I_T, I^{LV}_T) \leq M$, which exists because of (H1) and (H3). Let $\phi(z) = z^2$ on $[0,M+1]$ and $\phi(z) \in \mathcal{C}^{\infty}_b([0,\infty])$. We write
\[
V(t,x,i) = \E_{t,x}\left[ \phi \left(i+\int_t^T \lambda(u,\overline{X}_u) du\right)\right], \quad H = \frac{1}{2}(\partial^2_{xx}V- \partial_xV) + \partial_i V.
\]
 Consider the auxiliary diffusion
\[
d\hat{X}_u = (\lambda_x - \frac{1}{2}\lambda)(u, \hat{X}_u) du + \sqrt{\lambda(u,\hat{X}_u)}d\hat{B}_u
\]
such that $\hat{X}_t = x$. Then, following the same argument as in \cite{lucic2026gatheral} we have that
\[
H(t,x,i) =\hat{\E}_{t,x}\left[ e^{\int_t^T \varpi(u,\hat{X}_u)}\phi'\left( i + \int_t^T \lambda(u, \hat{X}_u)du\right)\right], \quad \varpi = \frac{1}{2}(\partial^2_{xx}\lambda- \partial_x\lambda).
\]
In the range where $\phi(z) = z^2$ we have
\[
H(t,x,i) = i \Theta(x,t) + g(t,x)
\]
where
\[
\Theta(t,x,) =2\hat{\E}_{t,x}\left[ e^{\int_t^T \varpi(u,\hat{X}_u) du}\right]
\]
\[
g(t,x) = 2\hat{\E}_{t,x} \left[e^{\int_t^T \varpi (u, \hat{X}_u)du}\int_t^T \lambda(u, \hat{X}_u ) du\right].
\]
This point is exactly the same as in Appendix C.1 of \cite{lucic2026gatheral}. We also have.
\[
\partial_t V + \lambda H = 0, \quad V(T,x,i) = \phi(i).
\]
The Itô formula yields
\begin{align*}
dV = &[\partial_t V + v_t (\frac{1}{2}(\partial^2_{xx}V-\partial_x V) + \partial_iV)] dt+ \rho \sqrt{v_t} \partial_x V dW_t \\
= & (\partial_t V + v_t H)dt + \rho \sqrt{v_t}\partial_x V dW_t \\
= & (v_t - \lambda(t,x))Hdt + \rho \sqrt{v_t}\partial_x V(t,x,i) dW_t.
\end{align*}
Since $V(T,X_T,I_T) = \phi(I_T)$ and $\psi(0,X_0, I_0) = \E[\phi(I_T^{LV})]$ we have
\[
\E[\phi(I_T)] - \E[\phi(I_T^{LV})] = \int_0^T \E\left[(v_t - \lambda(t,X_t))H(t,X_t,I_t) \right] dt.
\]
In the range where $\phi(z) = z^2$ we have
\begin{equation} \label{eq: Variance first reduction}
\Var(I_T) - \Var(I_T^{LV}) = \int_0^T \E[\eps_t H(t,X_t,I_t)]dt, \quad \eps_t = v_t - \lambda(t,X_t).
\end{equation}
Due to the fact that $\E[(v_t-\lambda(t,X_t)) f(X_t)] = 0$ for all $f(X_t) \in L^2(\Omega)$ we have that
\[
\E[\eps_t H(t,X_t,I_t)] = \E[\eps_tI_t \Theta(t,X_t)] + \E[\eps_t g(t,X_t)] =\E[\eps_tI_t \Theta(t,X_t)].
\]
The rest of the paper consists in using chaos expansions to deduce the asymptotic behavior of Equation \eqref{eq: Variance first reduction}.

\subsection{Asymptotic analysis}
The analysis of Equation \eqref{eq: Variance first reduction} is equivalent to the analysis of the integrand $\E[\eps_t H(t,X_t,I_t)]$. We rewrite it as
\[
\E[\eps_t H(t,X_t,I_t)] = \E[\eps_t I_t \Theta(t,X_t)] + \E[\eps_t g(t,X_t)] = \E[\eps_t I_t \Theta(t,X_t)],
\]
and we can further rewrite the previous expression as
\[
\E[\eps_t H(t,X_t,I_t)] =2 \E[\eps_t I_t] + \E[\eps_t(I_t - \E[I_t|X_t](\Theta(t,X_t) - 2)] = 2\E[\eps_t I_t] + \text{Error}(t).
\]
On the previous equation, the term $2\E[\eps_t I_t]$ is the one leading the dynamics of $\E[\eps_t H(t,X_t,I_t)]$. Thus, the asymptotic analysis consists in estimating the order of $\text{Error}(t)$ in $t$ and finding the leading contribution of $2\E[\eps_t I_t]$, since it is not directly computable in its current state.
\subsubsection{Analysis of the Error term}
We focus our effort in showing the following result.
\begin{prop} \label{prop: Error T^5}
    Under Hypotheses (H1) -- (H3) We have
    \[
    \int_0^T |\text{Error}(t)| dt = O(T^{5}).
    \]
\end{prop}
Observe that the error term in Proposition \ref{prop: Error T^5} can be studied using Cauchy-Schwarz inequality. Instead of applying the Cauchy-Schwarz inequality to the three terms
\[
\eps_t, \quad I_t - \E[I_t|X_t], \quad \Theta(t,X_t) - 2
\]
we will first work with an elementary bound for $\Theta(t,X_t)-2$.
\begin{lema}
    Under Hypotheses (H1)--(H3), there exists a constant $C > 0$ such that
    \[
    |\Theta(t,X_t) - 2|\leq C(T-t).
    \]
\end{lema} \label{lemma: estimate theta}
\begin{proof}
    Since $H = \frac{1}{2}$, the function $\varpi$ is bounded in. Therefore, there exists a generic constant $C > 0$ not depending on $T-t$ such that
    \[
    |\Theta(t,X_t) - 2 | \leq 2 |e^{||\varpi||_{\infty}(T-t)} -1| \leq C(T-t).
    \]
\end{proof}
We now apply Cauchy-Schwarz' inequality. Using Lemma \ref{lemma: estimate theta}, we can bound the error term as
\[
|\text{Error}(t)| \leq C (T-t) \E[|\eps_t(I_t-\E[I_t|X_t])|] \leq C(T-t)||\eps_t||_{L^2(\Omega)}||I_t-\E[I_t|X_t]||_{L^2(\Omega)}.
\]
The proof of Proposition \ref{prop: Error T^5} is achieved by estimating the norms
\[
||\eps_t||_{L^2(\Omega)}, \quad ||I_t- \E[I_t|X_t]||_{L^2(\Omega)}.
\]
\begin{lema}\label{lemma: bound I_t}
    Under Hypotheses (H1)--(H3) we have
    \[
    ||I_t - \E[I_t|X_t]||_{L^2(\Omega)} \leq Ct^{3/2}.
    \]
\end{lema}
\begin{proof}
    Minkowski's inequality allows us to bound $||I_t-\E[I_t|X_t]||_{L^2(\Omega)}$ as
    \[
    ||I_t - \E[I_t|X_t] ||_{L^2(\Omega)} \leq ||I_t - \E[I_t] ||_{L^2(\Omega)} \leq \int_0^t ||v_s - \E[v_s] ||_{L^2(\Omega)}.
    \]
    Then, using Proposition \ref{prop: v properties}-($v$-I) we have
    \[
    \int_0^t  ||v_s - \E[v_s] ||_{L^2(\Omega)} \leq \int_0^t ||v_s-v_0||_{L^2(\Omega)} ds \leq C t^{3/2},
    \]
    concluding the proof.
\end{proof}
The strategy for obtaining a bound for $||\eps_t||_{L^2(\Omega)}$ is again using the minimal norm property of the conditional expectation. Indeed, in order to bound $$||\eps_t||_{L^2(\Omega)} = ||v_t - \lambda(t,X_t)||_{L^2(\Omega)}$$ is enough to find a function $f(X_t)$ such that $||v_t-f(X_t)||_{L^2(\Omega)}$ is small. Then, using that
\[
||v_t-\lambda(t,X_t)||_{L^2(\Omega)} \leq ||v_t - f(X_t)||_{L^2(\Omega)}
\]
we will be able to find a bound for $||\eps_t||_{L^2(\Omega)}$.
\begin{lema} \label{lemma: error eps_t}
    Assume Hypotheses (H1)--(H3) hold. Let $X_t^c = X_t - \E[X_t]$ and let
    \[
    f(X_t) = \E[v_t] + \frac{\rho \beta_0}{\sigma_0}X_t^c + \frac{1}{v_0}\left(\frac{\beta_0\beta_0'}{2} - \frac{\beta_0^2}{4v_0} \right)\left( (X_t^c)^2 - \E[(X_t^c)^2]\right).
    \]
    Then,
    \[
    ||\eps_t||_{L^2(\Omega)} \leq ||v_t- f(X_t)||_{L^2(\Omega)} \leq Ct^{3/2}.
    \]
\end{lema}
\begin{proof}
    We resort to Proposition \ref{prop: v properties}- ($v$-III) and \ref{prop: X properties}- ($X$-II), ($X$-III) to say that processes $v$ and $X_t^c$ admit the following short-time representations:
    \begin{equation} \label{eq: expansion v}
        v_t= \E[v_t] + \beta_0W_t + \frac{\beta_0\beta_0'}{2}(W_t^2-t) + O_{L^2(\Omega)}(t^{3/2}),
    \end{equation}
    \begin{equation} \label{eq: expansion X}
        X_t^c = \rho \sigma_0 W_t + \frac{\rho \beta_0}{4\sigma_0}(W_t^2 -t) +O_{L^2(\Omega)}(t^{3/2}),
    \end{equation}
    and
    \begin{equation} \label{eq: expansion X^2}
        (X_t^c)^2 - \E[(X_t^c)^2] = v_0 (W_t^2 -t) + O_{L^2(\Omega)}(t^{3/2}).
    \end{equation}
    Then, straightforward computations show that
    \[
    v_t - f(X_t) = O(t^{3/2}),
    \]
    which concludes the result.
\end{proof}
Gathering toghether the two previous lemmas, we are able to prove Proposition \ref{prop: Error T^5}.
\begin{proof}{(of Proposition \ref{prop: Error T^5})}
    We aim to bound
    \[
    \int_0^T |\text{Error}(t)| dt.
    \]
    As mentioned before, Cauchy-Schwarz yields
    \[
    |\text{Error}(t)| \leq C (T-t) \E[\eps_t(I_t-\E[I_t|X_t])] \leq C(T-t)||\eps_t||_{L^2(\Omega)}||I_t-\E[I_t|X_t]||_{L^2(\Omega)}.
    \]
    Then, Lemma \ref{lemma: bound I_t} and Lemma \ref{lemma: error eps_t} allow us to derive the estimate
    \[
    |\text{Error}(t)| \leq C(T-t) t^3,
    \]
    for a generic constant $C > 0$ that does not depend on $T$ nor $t$. Integrating,
    \[
    \int_0^T |\text{Error}(t)|dt \leq C T^5,
    \]
    as desired.
\end{proof}
Proposition \ref{prop: Error T^5} allows us to write the difference of variances as
\[
\Var(I_T) - \Var(I_T^{LV}) = 2\int_0^T \E[\eps_tI_t] dt + O(T^5).
\]
Hence, in order to reach the conclusion in Theorem \ref{th: main}-(ii) we need to find the leading behavior of the term $2\int_0^T \E[\eps_tI_t] dt$. The next section is devoted to attain such goal.

\subsubsection{Analysis of the leading term}
Since $\eps_t$ is a centered random variable for all times $t$, we can write
\[
2\int_0^T \E[\eps_tI_t] dt = 2\int_0^T \Cov(\eps_t, I_t) dt.
\]
Using the chaos representation for $\eps_t$ and $I_t$, together with the representation \eqref{eq: stroock} we find
\begin{equation} \label{eq: cov as chaos}
\Cov(\eps_t, I_t) =  \int_0^t \E[D_r\eps_t]\E[D_rI_t] dr + R_t,
\end{equation}
where $R_t$ is a remainder term.

The decomposition of the covariance as the scalar product of Malliavin derivatives provided in \eqref{eq: cov as chaos} again simplifies our problem. Now, it is enough to find decompositions for each of the two terms. It is possible to find estimates for both $\E[D_r\eps_t]$ and $\E[D_rI_t]$, but the product does not give a satisfactory result. Instead, in order to find a better estimation, we will use a strategy similar as in Lemma \ref{lemma: error eps_t}. In spite of studying $\E[\eps_t I_t]$, we will analyze the equivalent quantity
\[
\Cov(\eps_t, I_t) =\E[\eps_t I_t] = \E[\eps_t (I_t - f(X_t)] = \Cov(\eps_t, I_t-f(X_t))
\]
so that we have
\[
\Cov(\eps_t, I_t-f(X_t)) = \int_0^t \E[D_r\eps_t]\E[D_r(I_t-f(X_t))]dr + R_t.
\]
The reason behind this decomposition lies on the fact that it is hard to obtain a sharp expansion for $\E[D_r\eps_t]$, in the sense that not all the lower order terms are easy to find explicitly, and the order of the higher-order terms is hard to identify without ambiguity. Nevertheless, we can compensate the lack of transparency in the expansion of $\E[D_r\eps_t]$ by expanding $\E[D_rI_t]$ up to high orders and then eliminating the lower order terms by choosing an appropriate function $f(X_t)$ using the same strategy as in Lemma \ref{lemma: error eps_t}. We first start analyzing $\E[D_r\eps_t]$.
\begin{prop} \label{prop: E(D_reps_t)}
    Under Hypotheses (H1)--(H3), there exists a function $P(t)$ with $|P(t)| \leq Ct^{1/2}$ such that
    \[
    \E[D_r\eps_t] =P(t) + \mathcal{B}r + O(t^{3/2}),
    \]
    where
    \[
    (\mathcal{L}f)(v) := \alpha(v)f'(v) + \frac{1}{2}\beta(v)^2 f''(v),
    \]
    \[
    \theta := \frac{\alpha_0}{2\sigma_0}- \frac{\beta_0^2}{8\sigma_0^3}, \quad \kappa := \frac{\beta_0\theta}{\sigma_0}+ \frac{\rho \beta_0^2}{2\sigma_0}, \quad \mathcal{B}:= (\mathcal{L} \beta)(v_0) - \alpha_0'\beta_0 - \kappa.
    \]
\end{prop}
\begin{proof}
    We write
   \begin{align}
       \E[D_r\eps_t] = &\E[D_r v_t] - \E\left[ \partial_x \lambda(t,X_t) D_rX_t\right] \\
       = & \E[D_rv_t] - \E[\partial_x \lambda(t,X_t)]\E[D_rX_t] - \Cov(\partial_x\lambda(t,X_t), D_rX_t). \label{eq: Dreps_t with covariance}
   \end{align}
   On the one hand, Proposition \ref{prop: Dv properties} -($Dv$-III) and Proposition \ref{prop: DX properties}-($DX$-III) we have that
   \[
   \E[D_rv_t] = \beta_0 + \alpha_0'\beta_0(t-r) + (\mathcal{L}\beta)_0 r + O(t^{3/2}),
   \]
   and
   \[
   D_rX_t - \E[D_rX_t] = \frac{\rho \beta_0}{2\sigma_0}W_t + R_{r,t}, \quad ||R_{r,t}||_{L^2(\Omega)} \leq Ct,
   \]
   where we have abbreviated $(\mathcal{L}\beta)_0 := (\mathcal{L}\beta)(v_0)$. Notice that
   \begin{align*}
   &\Cov(\partial_x\lambda(t,X_t), D_rX_t) \\
   = &\E\left[\left\{ \partial_x \lambda(t,X_t) - \E[\partial_x \lambda(t,X_t)]\right\}\left\{D_rX_t - \E[D_rX_t] \right\} \right] \\
   = & \frac{\rho \beta_0}{2\sigma_0}\E\left[\left\{ \partial_x \lambda(t,X_t) - \E[\partial_x \lambda(t,X_t)]\right\}\right] + \E\left[\left\{ \partial_x \lambda(t,X_t) - \E[\partial_x \lambda(t,X_t)]\right\}R_{r,t}\right]
   \end{align*}
   Let $\delta_t := \partial_x \lambda(t,X_t) - \E[\partial_x \lambda(t,X_t)]$, so we can write the previous equality as
   \[
   \Cov(\partial_x\lambda(t,X_t), D_rX_t) = \frac{\rho \beta_0}{2\sigma_0}\E[\delta_tW_t]  + \E[\delta_t R_{r,t}].
   \]
   Using Equation \eqref{eq: Poincare}, we can estimate $||\delta_t||_{L^2(\Omega)}$ as
   \[
   ||\delta_t||_{L^2(\Omega)}^2 =  \Var(\delta_t) \leq \E\left[\int_0^t |\partial^2_{xx}(t,X_t)|^2 |D_uX_t|^2 du \right] \leq Ct
   \]
   because both $\partial^2_{xx} \lambda(t,x)$ and $\sup_{u \leq t} \E[|D_u X_t|^2]$ are uniformly bounded. 
   
   We finally deal with $\E[\partial_x\lambda(t,X_t)]$. First, notice that by definition of $\lambda$ the following equality holds
   \[
   \E[v_t X_t^c] = \E[\lambda(t,X_t) X_t^c].
   \]
   A Taylor expansion of $\lambda(t,X_t)$ around the point $\overline{x} = \E[X_t]$ allows us to write
   \[
   \E[\lambda(t,X_t)X_t^c] = \partial_x \lambda(t,\overline{x})\E[(X_t^c)^2] + O(t^{3/2})
   \]
   because $\E[(X_t^c)^3] = O(t^{3/2})$. Now, using Proposition \ref{prop: v properties} - ($v$-III) and \ref{prop: X properties}-($X$-II) we have, due to the orthogonality of $W_t$ and $W_t^2-t$,
   \[
   \E[v_tX_t^c] = \rho \sigma_0\beta_0t + O(t^2).
   \]
   Using Proposition \ref{prop: X properties}-($X$-IV) and isolating $\partial_x\lambda(t,\overline{x})$ we find that
   \[
   \partial_x\lambda(t,\overline{x}) = \frac{\rho \beta_0}{\sigma_0} + O(t^{1/2}).
   \]
   Finally, since the mean value theorem yields
   \[
   \E\left[|\partial_x\lambda(t,\overline{x}) - \partial_x\lambda(t,X_t)| \right] \leq C \sqrt{\Var(X_t)} \leq Ct^{1/2}
   \]
   we have that there exists $\mathcal{E}(t)$ with $|\mathcal{E}(t)|\leq Ct^{1/2}$ such that
   \[
    \E[\partial_x\lambda(t,X_t)] = \frac{\rho \beta_0}{\sigma_0} + \mathcal{E}(t).
   \]
   Going back to Equation \eqref{eq: Dreps_t with covariance}, we have that
   \begin{align*}
   &\E[D_r\eps_t]\\
   =& \left(\alpha'_0 \beta_0 + \frac{\rho \beta_0^2}{2\sigma_0} \right)t - \rho\sigma_0 \mathcal{E}(t) - \frac{\rho \beta_0}{2\sigma_0}\E[\delta_tW_t] + \left( (\mathcal{L}\beta)_0 - \alpha'_0\beta_0 - \left( \frac{\beta_0\theta}{\sigma_0}+ \frac{\rho \beta_0^2}{2\sigma_0}\right)\right)r + O(t^{3/2})
   \end{align*}
   so the proof is complete with
   \[
   P(t):= \left(\alpha'_0 \beta_0 + \frac{\rho \beta_0^2}{2\sigma_0} \right)t - \rho\sigma_0 \mathcal{E}(t) - \frac{\rho \beta_0}{2\sigma_0}\E[\delta_tW_t].
   \]
\end{proof}
The previously mentioned lack of transparency in the expansion of $\E[D_r\eps_t]$ is encapsulated in the function $P(t)$, whose value is not explicit. Luckily, we can find $f(X_t)$ such that $\E[D_r(I_t - f(X_t))]$ is orthogonal to $P(t)$ in $L^2(0,t)$ and it provides the leading asymptotic order of $\Cov(\eps_t, I_t)$. 
\begin{prop} \label{prop: covariance in t}
    Let $\mathcal{B}$ be defined as in Proposition \ref{prop: E(D_reps_t)}. Under Hypotheses (H1)--(H3) we have
    \[
    \Cov(\eps_t, I_t) = -\frac{\mathcal{B}\beta_0}{12}t^3 + O(t^{7/2}).
    \]
\end{prop}
\begin{proof}
   Let
   \[
   f(X_t) = \E[I_t] + \frac{\rho \beta_0 t}{2\sigma_0}X_t^c
   \]
   and define $\mathcal{I}_t =  I_t - f(X_t) = I_t - \E[I_t] - \frac{\rho \beta_0 t}{2\sigma_0}X_t^c$. As previously mentioned, we write
   \[
   \Cov(\eps_t, I_t) = \Cov(\eps_t, \mathcal{I}_t) = \int_0^t \E[D_r\eps_t] \E[D_r\mathcal{I}_t] dr + R_t
   \]
   with
   \[
   |R_t|\leq ||\eps_t^{\geq 2}||_{L^2(\Omega)}||\mathcal{I}_t^{\geq 2}||_{L^2(\Omega)}.
   \]
   On the one hand, we have due to Lemma \ref{lemma: error eps_t},
   \[
   ||\eps^{\geq 2}_t||_{L^2(\Omega)} \leq ||\eps_t||_{L^2(\Omega)} \leq Ct^{3/2}.
   \]
   On the other hand, by definition of $\mathcal{I}_t$ and the expansions given in Proposition \ref{prop: v properties}-($v$-III) and \ref{prop: X properties}-($X$-II) we have
   \[
   ||\mathcal{I}_t^{\geq 2}||_{L^2(\Omega)} \leq ||I_t^{\geq 2}||_{L^2(\Omega)} + C|t| \cdot ||(X_t^c)^{\geq 2}||_{L^2(\Omega)} \leq Ct^2.
   \]
   Hence, the Cauchy-Schwarz inequality yields
   \[
   |R_t|\leq Ct^{7/2}.
   \]
   Regarding $\E[D_r\mathcal{I}_t]$, we resort to the fact that
   \[
   \E[D_rI_t] = \beta_0(t-r) + O(t^2),
   \]
   in order to obtain
   \[
   \E[D_r\mathcal{I}_t] = \E[D_rI_t]- \frac{\rho \beta_0 t}{2\sigma_0}\E[D_rX_t^c] =    \beta_0\left( \frac{t}{2}-r\right) + O(t^2).
   \]
   Notice that
   \[
   \int_0^t P(t)\left(\frac{t}{2}-r \right)dr =0
   \]
   for any function $P(t)$.  Plugging the expansions into de covariance equation, we have
   \begin{align*}
   \int_0^t \E[D_r\eps_t]\E[D_r\mathcal{I}_t]dr = &\int_0^t \beta_0\left(\frac{t}{2}-r \right)(P(t)+\mathcal{B}r + O(t^{3/2})) dr \\
   =& T_1 + T_2  + O(t^{7/2})
   \end{align*}
   where
   \[
   T_1 = \beta_0P(t) \int_0^t \left(\frac{t}{2}-r\right)dr, \quad T_2 = \beta_0\mathcal{B}\int_0^t r\left(\frac{t}{2}-r \right)dr,
    \]
   Integrating, we obtain $T_1 =0$ and 
   \[
   T_2 = -\frac{\mathcal{B}\beta_0}{12}t^3.
   \]
   Thus,
   \[
   \Cov(\eps_t,I_t) = -\frac{\mathcal{B}\beta_0}{12}t^3 + O(t^{7/2}).
   \]
    \end{proof}
   This allows us to prove Theorem \ref{th: main}--(ii).
   \begin{proof}{(of Theorem \ref{th: main}-(ii))}
       We have
       \[
       \Var(I_T) - \Var(I_T^{LV}) = 2\int_0^T \E[\eps_tI_t] dt + O(T^5) = 2\int_0^T \Cov(\eps_t, \mathcal{I}_t) dt + O(T^5).
       \]
       Then, using Proposition \ref{prop: covariance in t} we are lead to
       \[
        \Var(I_T) - \Var(I_T^{LV}) = - \frac{\mathcal{B}\beta_0}{24}T^4 + O(T^{9/2}).
       \]
       Using the definition of $\mathcal{B}$ we see that 
       \[
     \Var(I_T) - \Var(I_T^{LV}) = \frac{1}{24}\left\{\alpha'_0 \beta_0^2+ \frac{\alpha_0\beta_0^2}{2v_0} - \frac{\beta_0^4}{8v_0^2} + \frac{\rho \beta_0^3}{2\sigma_0}-\alpha_0 \beta_0 \beta_0' - \frac{\beta_0^3\beta_0^{''}}{2} \right\}T^4 + O(T^{9/2}).
       \]
       Defining $K_0(\alpha,\beta,\rho)$ as
       \[
       K_0(\alpha,\beta,\rho) =\alpha'_0 \beta_0^2+ \frac{\alpha_0\beta_0^2}{2v_0} - \frac{\beta_0^4}{8v_0^2} + \frac{\rho \beta_0^3}{2\sigma_0}-\alpha_0 \beta_0 \beta_0' - \frac{\beta_0^3\beta_0^{''}}{2}
       \]
       we conclude the result.
   \end{proof}

\section{Undoing the localization argument and examples} \label{sec: loc and examples}
In the remaining of the paper, we want to apply Theorem \ref{th: main} to well-known examples of the literature in order to see how the refutation conditions look like in those models. The vast majority of classical stochastic volatility models, however, do not fit into Hypotheses (H1)--(H3). Providing examples is therefore a two-step procedure. First, we show how we can relax (H1)--(H3) and then we will compute the refutation conditions for particular models.

\textbf{(H'1)} The variance $v$ is of the form \eqref{eq: variance model} with $\alpha \in \mathcal{C}^1(U)$, and $\beta \in \mathcal{C}^2(U)$ where $U:= (v_0 - r, v_0 + r)$ and, moreover, $\sup_{t \in [0,T]}||v_t||_{L^p(\Omega)} < \infty$ for all $p \geq 2$.

To see that the result can be extended to (H'1), let $\alpha_n$, $\beta_n$ be smooth regularized versions of $\alpha$ and $\beta$ such that $\alpha_n = \alpha$ and $\beta_n = \beta$ on $U = (v_0 - r, v_0 + r)$ and $\alpha_n, \beta_n \equiv 0$ outside of $U_n := (v_0-r-\frac{1}{n}, v_0 - r+\frac{1}{n})$. Such regularization can be constructed by considering $\alpha_n(x) = \chi_n(x)\alpha(x)$ where $\chi_n$ denotes a smooth bump function. Consider the auxiliary family SDE parametrized by $\lambda >0$,
\begin{equation} \label{eq: auxiliary}
    dY_t^{\lambda} = \lambda\alpha_n(Y^{\lambda}_t) dt + \lambda\beta_n(Y^{\lambda}_t)dW_t.
\end{equation}
Since, by construction, the sample paths of $Y_t^{\lambda}$ belong to $\overline{U}_n = [v_0-r-\frac{1}{n}, v_0 + r + \frac{1}{n}]$ then by \cite[Theorem 3.5]{alfonsi2025nonnegativity} the ample paths of the solution $v^n$ to the Volterra SDE
\[
v_t^n = v_0 + \int_0^t K(t-s)\alpha_n(v^n_s) ds + \int_0^t K(t-s)\beta_n(v_s^n) dW_s
\]
also belong to $\overline{U}_n$, so Hypotheses (H1) and (H2) are satisfied. Let $v^n$ and $\lambda_n(t,x)$ denote the stochastic and local variance respectively for the regularized model. Let $\overline{X}_t^n$ denote the log-price under the projection of the regularized variance $v_t^n$. Regarding (H3), it is not clear whether $\lambda_n(t,x)$ satisfies the bound in (H3). To avoid checking it, consider another localization. The localization strategy will differ in the case $H < \frac{1}{2}$ and the case $H = \frac{1}{2}$.

\subsection{Undoing the localization: Rough case}
For $H < \frac{1}{2}$, let $K^* = \E[I_T]$ and let $K^*_n = \E[I_T^n]$ where
\[
I_T = \int_0^T v_t dt, \quad I_T^n = \int_0^T v_t^n dt.
\]
In order to avoid checking (H3) in this case, we regularize again the local variance function. We sketch the localization argument for the sake of conciseness. Let $m^n_t = \E[v_t]$, $m^{X,n}_t = \E[X^n_t]$ and let
\[
c_H = \frac{\rho \beta_0}{\sigma_0 \Gamma(H+3/2)}.
\]
Let $\eta$ denote a smooth compactly supported function such that $\eta(x) \equiv 1$ in a small neighborhood around the origin. Define
\[
a_n(t,x) = v_0 + \eta(t^{H-\frac{1}{2}}(x-m^n_t))\left( m^{X,n}_t -v_0 + c_Ht^{H-\frac{1}{2}}(x-m_t^n)\right).
\]
Then, for sufficiently small $t > 0$ we have
\[
\sup_{x \in \R} |\partial_x^k a_n(t,x)| \leq C_n t^{k(H-\frac{1}{2})}, \quad ||\lambda_n(t,\overline{X}_t^n) - a_n(t,\overline{X}_t^n)||_{L^2(\Omega)} \leq C t^{2H}.
\]
 We also define
 \[
 J_T^n = \int_0^T a(t,\overline{X}_t^n) dt.
 \]
The function $a_n(t,x)$ satisfies (H3) and
 \[
 ||I_T^{LV,n} - J_T^n ||_{L^2(\Omega)} = O(T^{2H+1}).
 \]
 Moreover, it can be checked that the leading behavior of the ATM-IV level with underlying asset $J_T^n$ coincides with the one with $I_T^{LV,n}$ as underlying. Hence, for the purpose of deriving the leading ATM asymptotics, we may assume without loss of generality that the local variance coefficient used in the integrated functional satisfies (H3), replacing $\lambda_n$ by $a_n$ whenever necessary. We write
\begin{align*}
\E[(I_T-K^*)_+] - \E[(I_T^{LV}-K^*)] = \mathcal{E}_1 + \E[(I_T^n - K^*_n)_+] - \E[(I_T^{LV, n}-K^*_n)_+] + \mathcal{E}_2,
\end{align*}
with
\[
\mathcal{E}_1 = \E[(I_T - K^*)_+]- \E[(I_T^n - K^*_n)_+],
\]
\[
\mathcal{E}_2 = \E[(I_T^{LV,n} - K^*_n)_+]-\E[(I^{LV}_T - K^*)_+].
\]

Observe that, in the ATM regime, we can write
\[
\E[(I^n_T - K^*_n)_+] = K^*_n \left(2\Phi\left(\frac{IV_n(T,K^*_n)\sqrt{T}}{2} \right)-1 \right).
\]
Therefore, since
\[
\lim_{x \to 0}\frac{2\Phi(x/2)-1}{x} = \frac{1}{\sqrt{2\pi}}
\]
we have
\[
\lim_{T \to 0} \frac{\E[(I^n_T - K^*_n)_+]}{T^{H+1}} = \frac{v_0}{\sqrt{2\pi}}\lim_{T \to 0}\frac{IV_n(T,K^*)}{T^{H-\frac{1}{2}}}.
\]
Therefore,
\[
\lim_{T \to 0}\frac{ \E[(I_T^n - K^*_n)_+] - \E[(I_T^{LV, n}-K^*_n)_+]}{T^{H+1}} = \frac{v_0}{\sqrt{2\pi}}(\ell - \ell^{LV}) < 0.
\]
The objective in order to successfully transfer the asymptotic behavior we need to show that
\[
\lim_{T \to 0} \frac{|\mathcal{E}_1 + \mathcal{E}_2|}{T^{H+1}} = 0. 
\]
We first analyze $\mathcal{E}_1$. Since the function $(x-k)_+$ is Lipschitz in both variables $x$ and $k$, we have 
\[
|\mathcal{E}_1| \leq ||I_T-I_T^n||_{L^1(\Omega)} + |K^* - K^*_n| \leq 2||I_T- I_T^n||_{L^1(\Omega)}
\]
because $|K^* - K^*_n | = |\E[I_T] - \E[I_T^n]| \leq \E[|I_T-I_T^n|]$. Define the stopping time
\[
\tau_n = \inf\{t \geq 0 ; v_t \neq v_t^n \}.
\]
Then,
\[
|I_T - I_T^n| = |I_T - I_T^n|\1_{\{\tau_n \leq T\}}
\]
and therefore, Cauchy-Schwarz shows that
\[
||I_T-I_T||_{L^1(\Omega)} \leq (||I_T||_{L^2(\Omega)} + ||I_T^n ||_{L^2(\Omega)} )\sqrt{\P(\tau_n \leq T)} \leq C_n \sqrt{\P(\tau_n \leq T)}
\]
since $||I_T||_{L^2(\Omega)} + ||I_T^n ||_{L^2(\Omega)}$ can be bounded by a constant that does not depend on $T$. In order to analyse the stopping time $\tau_n$, we borrow an estimation from \cite{bures2026skew}. In the cited article, the authors provide quantitative bounds for $\tau_n$ in the case of the rough Heston model, but the same computations work for arbitrary functions $\alpha$ and $\beta$ satisfying (H'1). 
\begin{lema} \label{lemma: bound stopping time}
    Under (H'1), there exists a constant $\delta_n$ such that for all $p \geq 2$, there exists $C_p > 0$ with
    \[
    \P(\tau_n \leq T) \leq C_p \delta_n^{-p} T^{pH}.
    \]
\end{lema}
\begin{proof}
    Follows the same strategy as in \cite{bures2026skew}- Proposition 5.4.
\end{proof}
Notice that we therefore have that there exists a generic constant $C_{p,n,H} > 0$ that does not depend on $T$ such that
\[
\frac{|\mathcal{E}_1|}{T^{H+1}} \leq C_{p,n,H} T^{-1+\left( \frac{p}{2}-1\right)H}
\]
which tends to $0$ as $T \to 0$ if $p \geq 2$ is large enough. For the local variance model, notice that the estimate
\begin{align*}
|\mathcal{E}_2| \leq & ||I_T^{LV}-I_T^{LV,n}||_{L^1(\Omega)} + |K^* - K^*_n| \\
\leq &||I_T^{LV} - I_T^{LV,n}||_{L^1(\Omega)} + ||I_T - I^n_T||_{L^1(\Omega)}
\end{align*}
holds. In contrast with $||I_T-I_T^n||_{L^1(\Omega)}$, the analysis of the integrated local variances is more involved. We first have
\[
||I_T^{LV}-I_T^{LV,n} ||_{L^1(\Omega)}= \E\left[ \left|\int_0^T (\lambda(t,\overline{X}_t) - \lambda_n(t,\overline{X}_t^n))dt \right| \right],
\]
We define
\[
\Lambda(t):= \int_0^t ||\lambda(s,\overline{X}_s) - \lambda_n(s,\overline{X}^n_s) ||_{L^1(\Omega)} ds.
\]
Notice that the interest in the function $\Lambda$ resides on the fact that
\[
||I_T^{LV}-I_T^{LV,n} ||_{L^1(\Omega)} \leq \Lambda(T).
\]
The following bound for $\Lambda$ holds.
\begin{align*}
    \Lambda(t) \leq & \int_0^t ||(\lambda-\lambda_n)(s,\overline{X}_s)||_{L^1(\Omega)}ds + C_n\int_0^t s^{H-\frac{1}{2}}||\overline{X}_s - \overline{X}_s^n||_{L^1(\Omega)}ds.
    \end{align*}
Define $A(t) := \int_0^t ||(\lambda-\lambda_n (s,\overline{X}_s)||_{L^1(\Omega)}ds$. Observe that we can bound $A(t)$ in terms of $\P(\tau_n \leq t)$ in the following way.
\begin{lema} \label{lema: bound A_(t)}
    Under Hypothesis (H'1), there exists a generic constant $C > 0$ such that
    \[
    A(t) \leq C t^{1+\frac{pH}{2}}.
    \]
\end{lema}
\begin{proof}
Since $\overline{X}_s$ and $X_s$ have the same law, we can write
\[
A(t)=\int_0^t ||(\lambda-\lambda_n)(s,X_s)||_{L^1(\Omega)}\ ds.
\]
We let $\mathcal{A}_n :=\{\tau_n \leq t\}$ and define the auxiliary function $E_s(x) = \E[v_s | X_s = x, \mathcal{A}_n^c]$ for $s \leq t$. Notice that
\[
E_s(x) = \E[v_s | X_s = x, \mathcal{A}_n^c] = \E[v^n_s | X^n_s = x, \mathcal{A}_n^c].
\]
By the definition of conditional expectation with respect to an event we have
\[
E_s(x) = \frac{\E[v_s\1_{\mathcal{A}_n^c}| X_s = x]}{\P(\mathcal{A}_n^c |X_s = x)}.
\]
We split $A(t)$ into the terms
\[
A(t) \leq \int_0^t||\lambda(s,X_s) - E_s(X_s) ||_{L^1(\Omega)}ds + \int_0^t||\lambda_n(s,X_s) - E_s(X_s) ||_{L^1(\Omega)}ds.
\]
On the one hand
\begin{align*}
    \lambda(s,x) - E_s(x) = & \E[v_s\1_{\mathcal{A}_n} |X_s = x] +\E[v_s\1_{\mathcal{A}_n^c} | X_s = x ] - \frac{\E[v_s\1_{\mathcal{A}_n^c}| X_s = x]}{\P(\mathcal{A}_n^c |X_s = x)} \\
    = &\E[v_s\1_{\mathcal{A}_n} |X_s = x]  + \frac{\E[v_s\1_{\mathcal{A}_n^c}| X_s = x](\P(\mathcal{A}_n^c|X_s=x) - 1)}{\P(\mathcal{A}_n^c |X_s = x)} \\
    = &\E[v_s\1_{\mathcal{A}_n} |X_s = x]  - E_s(x) \P(\mathcal{A}_n |X_s = x) \\
    = & \E[(v_s - E_s(x)) \1_{\mathcal{A}_n} | X_s = x].
\end{align*}
and therefore
\[
||\lambda(s,X_s) - E_s(X_s)||_{L^1(\Omega)} \leq ||v_s\1_{\mathcal{A}_n}||_{L^1(\Omega)} + C \P(\mathcal{A}_n) \leq C \sqrt{\P(\mathcal{A}_n)}\left(||v_s||_{L^2(\Omega)} + \sqrt{\P(\mathcal{A}_n)}\right)
\]
which implies that
\[
||\lambda(s,X_s) - E_s(X_s) ||_{L^1(\Omega)} \leq C t^{\frac{pH}{2}}.
\]
Regarding the second integral, we use that $X_s = X_s^n$ on $\mathcal{A}_n^c$ to write
\begin{align*}
    ||\lambda_n(s,X_s) - E_s(X_s) ||_{L^1(\Omega)} = &  \E\left[ |\lambda_n(s,X_s) - E_s(X_s) | \1_{\mathcal{A}_n}\right] +  \E\left[ |\lambda_n(s,X_s) - E_s(X_s) | \1_{\mathcal{A}^c_n}\right] \\
    = &\E\left[ |\lambda_n(s,X_s) - E_s(X_s) | \1_{\mathcal{A}_n}\right] +  \E\left[ |\lambda_n(s,X^n_s) - E_s(X^n_s) | \1_{\mathcal{A}^c_n}\right] \\
     \leq &\E\left[ |\lambda_n(s,X_s) - E_s(X_s) | \1_{\mathcal{A}_n}\right] + ||\lambda_n(s,X_s^n) - E_s(X_s^n)||_{L^1(\Omega)}.
\end{align*}
Using the same approach as before,
\[
||\lambda_n(s,X_s^n) - E_s(X_s^n)||_{L^1(\Omega)} = \E\left[\left|\E[(v_s^n - E_s(X_s^n))\1_{\mathcal{A}_n}|X_s^n]\right| \right]\leq  C\P(\mathcal{A}_n).
\]
Then, using that both $\lambda_n$ and $E_s(x)$ are bounded we have
\[
\E\left[ |\lambda_n(s,X_s) - E_s(X_s) | \1_{\mathcal{A}_n}\right] \leq C\P(\mathcal{A}_n).
\]
Joining the estimations, we have that there exists a generic constant $C > 0$ such that
\[
A(t) \leq C\int_0^t \sqrt{\P(\tau_n \leq t)}\ ds \leq Ct^{1+\frac{pH}{2}},
\]
which concludes the desired result.
\end{proof}
For the second term we won't get a quantitative bound. In fact, we will reproduce the initial quantity we desire to bound in order to conclude the result using a Gronwall argument.
\begin{lema} \label{lema: bound second term}
    Under Hypothesis (H'1) we have
    \[
    \int_0^t s^{H-\frac{1}{2}}||\overline{X}_s - \overline{X}_s^n ||_{L^1(\Omega)} \leq \int_0^t s^{H-\frac{1}{2}}\sqrt{\Lambda(s)}ds.
    \]
\end{lema}
\begin{proof}
    We write the difference $\overline{X}_t-\overline{X}_t^n$ as
    \begin{align*}
        \overline{X}_s- \overline{X}_s^n = &-\frac{1}{2}\int_0^s (\lambda(u,\overline{X}_u) - \lambda_n(u,\overline{X}_u^n)) du \\
        + & \int_0^s (\hat{\sigma}(u, \overline{X}_u) - \hat{\sigma}(u,\overline{X}_u^n) ) d\hat{W}_u,
    \end{align*}
    where we recall that $\hat{\sigma}(t,x)=  \sqrt{\lambda(t,x)}$. For the drift term, we use Minkowski's inequality to say that
    \[
    ||\int_0^s (\lambda(u,\overline{X}_u) - \lambda_n(u,\overline{X}_u^n)) du ||_{L^1(\Omega)} \leq \int_0^s ||\lambda(u,\overline{X}_u) - \lambda_n(u,\overline{X}_u^n)||_{L^1(\Omega)} du = \Lambda(s).
    \]
    For the diffusion term, we have
    \[
   || \int_0^s (\hat{\sigma}(u, \overline{X}_u) - \hat{\sigma}(u,\overline{X}_u^n) ) d\hat{W}_u ||_{L^1(\Omega)} \leq   || \int_0^s (\hat{\sigma}(u, \overline{X}_u) - \hat{\sigma}(u,\overline{X}_u^n) ) d\hat{W}_u ||_{L^2(\Omega)}.
    \]
    The Itô isometry together with the elementary inequality $|\sqrt{a}-\sqrt{b}|^2 \leq |a-b|$ allows us to estimate
    \[
     || \int_0^s (\hat{\sigma}(u, \overline{X}_u) - \hat{\sigma}_n(u,\overline{X}_u^n) ) d\hat{W}_u ||_{L^2(\Omega)}^2 \leq \int_0^s ||\lambda(u,\overline{X}_u) - \lambda_n(u,\overline{X}_u^n) ||_{L^1(\Omega)} du = \Lambda(s),
    \]
    so 
    \[
    || \int_0^s (\hat{\sigma}(u, \overline{X}_u) - \hat{\sigma}(u,\overline{X}_u^n) ) d\hat{W}_u ||_{L^1(\Omega)} \leq \sqrt{\Lambda(s)}.
    \]
    The estimates for the drift and diffusion yield
    \[
    ||\overline{X}_s- \overline{X}_s^n||_{L^1(\Omega)} \leq \frac{1}{2}\Lambda(s) + \sqrt{\Lambda(s)}.
    \]
    Finally, using that
    \[
    \Lambda(s) \leq \int_0^s (\E[v_u] + \E[v^n_u]) du \leq C
    \]
    for $s$ small enough we have that
    \[
    ||\overline{X}_s- \overline{X}_s^n||_{L^1(\Omega)} \leq C \sqrt{\Lambda(s)}.
    \]
    This concludes the proof.
\end{proof}
Lemmas \ref{lema: bound A_(t)} and \ref{lema: bound second term} allow us to write
\[
\Lambda(t) \leq C t^{1+\frac{pH}{2}} + \int_0^t s^{H-\frac{1}{2}}\sqrt{\Lambda(s)}ds.
\]
Hence, by the Gronwall lemma in \cite{ou1957boundedness} and \cite{cheung2006discrete} we have that
\[
\Lambda(t) \leq C \left( T^{1+\frac{pH}{2}}  + T^{1+2H}\right).
\]
In particular,
\[
\lim_{T \to 0} \frac{|\mathcal{E}_2|}{T^{H+1}} \leq \lim_{T \to 0} \frac{\Lambda(T)}{T^{H+1}} + \frac{||I_T-I_T^n||_{L^1(\Omega)}}{T^{H+1}} = 0.
\]
This concludes that for a variance process $v$ satisfying (H'1), we have
\[
\lim_{T \to 0} \frac{\E[(I_T-K^*)] - \E[(I_T^n - K^*)_+]}{T^{H+1}} < 0
\]
for 
\[
\rho^2 > \frac{2H+3}{4},
\]
extending Theorem \ref{th: main}-(i) for weaker hypotheses on $\alpha$ and $\beta$.
\subsection{Undoing the localization: Markovian case}
We now let $v$ follow Hypothesis (H'1) with $H = \frac{1}{2}$ and we consider the same regularization $v^n$ of $v$ as with the rough case.
For $H=\frac12$ and $|\rho|=1$, we introduce a smooth approximation of the projected local variance $\lambda_n$. Again, we sketch the argument for the sake of conciseness. Define, in a neighborhood of $v_0$,
\[
d(v)=\frac{\rho\beta(v)}{\sqrt v},
\qquad
h(v)=\frac{\alpha(v)}{d(v)}+\frac v2-\frac v2d'(v),
\]
and let
\[
\begin{aligned}
P_3(t,z)
={}&v_0+d_0z+\frac{d_0d'_0}{2}z^2
+\frac{d_0((d'_0)^2+d_0d''_0)}{6}z^3\\
&+d_0h_0t+\left(d_0d'_0h_0+\frac{d_0^2h'_0}{2}\right)tz.
\end{aligned}
\]
These coefficients are unchanged by localization, since $\alpha_n,\beta_n$ agree with $\alpha,\beta$ near $v_0$. Let $\eta$ be a smooth compactly supported function satisfying $0\le\eta\le1$ and equal to one near the origin. Define
\[
a_n(t,x)=v_0+\eta(x-X_0)\bigl(P_3(t,x-X_0)-v_0\bigr).
\]
It can be shown that
\[
\sup_{x \in \R} |\partial_x^k a_n(t,x)| \leq C_{n,k}, \quad ||\lambda_n(t,\overline{X}_t^n)- a_n(t,\overline{X}_t^n)||_{L^2(\Omega)} \leq C_n t^2.
\]
Thus, if we let $J_T^n = \int_0^T a_n(t,\overline{X}^n_t)dt$ one can check that
\[
||I_T^{LV,n}-J_T^n||_{L^2(\Omega)} = O(T^3),
\]
which implies that
\[
\Var(I_T^{LV,n}) - \Var(J_T^n) = O(T^{9/4}).
\]
Hence, for the purpose of computing the $T^4$ coefficient in the difference of variances between the original model integrated variance and its integrated local variance, we can replace $\lambda_n$ by $a_n$ when needed since this regularization preserves the leading coefficient. We again let
\[
\tau_n = \inf\{t \geq 0 ; v_t \neq v_t^n\}.
\]
The strategy follows the same lines as with the rough case, but we will rely on the fact that $H = \frac{1}{2}$ in order to invoke some shortcuts. Recall that, under the localized model,
\[
\lim_{T \to 0} \frac{\E[(I_T^n)^2] - \E[(I_T^{LV,n})^2]}{T^4} < 0.
\]
Thus, in order to show that
\[
\lim_{T \to 0} \frac{\E[(I_T)^2] - \E[(I_T^{LV})^2]}{T^4} < 0.
\]
it is enough to show that
\[
\lim_{T \to 0} \frac{(\E[(I_T)^2] - \E[(I_T^n)^2]) - (\E[(I_T^{LV})^2] - \E[(I_T^{LV,n})^2]) }{T^4} = 0.
\]
On the one hand,
\[
\E[(I_T)^2] - \E[(I_T^n)^2] = \E\left[ \left\{(I_T)^2 - (I_T^n)^2 \right\} \1_{\{\tau_n \leq T\}}\right] \leq C \sqrt{\P(\tau_n \leq T)}
\]
Using the bound on Lemma \ref{lemma: bound stopping time} we can choose $p \geq 2$ large enough so that
\[
\lim_{T \to 0} \frac{\sqrt{\P(\tau_n \leq T)}}{T^4} = 0
\]
and therefore
\[
\lim_{T \to 0} \frac{\E[(I_T)^2] - \E[(I_T^n)^2]}{T^4} = 0.
\]
For the local variances, we use Cauchy-Schwarz in order to say that
\[
|\E[(I_T^{LV})^2] - \E[(I_T^{LV,n})^2]| \leq ||I_T^{LV} -I_T^{LV,n}||_ {L^2(\Omega)}(||I_T^{LV}||_{L^2(\Omega)} + ||I_T^{LV,n}||_{L^2(\Omega)}).
\]
It is therefore enough to bound  $||I_T^{LV} -I_T^{LV,n}||_ {L^2(\Omega)}.$ The strategy is similar as the one in the rough case, we first split the difference into two terms,
\[
|I_T^{LV} -I_T^{LV,n}||_ {L^2(\Omega)} \leq A(T) + C\int_0^T ||\overline{X}_t - \overline{X}_t^n||_{L^2(\Omega)}
\]
where
\[
A(t):= \int_0^t ||(\lambda-\lambda_n)(s,\overline{X}_s) ||_{L^2(\Omega)}.
\]
The same argument as with the rough case yields
\begin{equation} \label{eq: bound lambda vs lambda_n}
||(\lambda-\lambda_n)(s,\overline{X}_s) ||_{L^2(\Omega)} \leq C\P(\tau_n \leq T)^{1/4}.
\end{equation}
We skip it for the sake of conciseness. Regarding $ ||\overline{X}_t - \overline{X}_t^n||_{L^2(\Omega)}$, we write
\begin{align*}
    \overline{X}_t- \overline{X}_t^n = -\frac{1}{2}&\int_0^t \lambda(s, \overline{X}_s) - \lambda_n(s,\overline{X}^n_s) ds \\
    + & \int_0^t \hat{\sigma}(s,\overline{X}_s) -\hat{\sigma}_n(s,\overline{X}^n_s) d\hat{W}_s.
\end{align*}
Regarding the drift term, using the bound from Equation \eqref{eq: bound lambda vs lambda_n} we obtain
\begin{align*}
&||\frac{1}{2}\int_0^t \lambda(s, \overline{X}_s) - \lambda_n(s,\overline{X}^n_s) ds ||_{L^2(\Omega)} \\
\leq &\frac{t}{2}\sup_{t \in [0,T]} || \lambda(t, \overline{X}_t) - \lambda_n(t,\overline{X}^n_t)  ||_{L^2(\Omega)} \\
\leq & \frac{t}{2} \left( \sup_{t \in [0,T]} ||(\lambda-\lambda_n)(t,\overline{X}_t) ||_{L^2(\Omega)} +\sup_{t \in [0,T]} || \lambda_n(t, \overline{X}_t) - \lambda_n(t,\overline{X}^n_t)  ||_{L^2(\Omega)} \right) \\
\leq &C \frac{t}{2} \left(  \P(\tau_n \leq T)^{1/4} + \sup_{t \in [0,T]} ||\overline{X}_t-\overline{X}_t^n||_{L^2(\Omega)}\right).
\end{align*}
For the diffusion term, using Burkholder-Davis-Gundy inequality, the fact that
\[
|\hat{\sigma}(s,\overline{X}_s) - \hat{\sigma}_n(s,\overline{X}^n_s)| \leq C|\lambda(s,\overline{X}_s) - \lambda_n(s,\overline{X}^n_s)|
\]
and a similar strategy of splitting the difference between $\lambda$ and $\lambda_n$ we find
\begin{align*}
||\int_0^t \hat{\sigma}(s,\overline{X}_s) -\hat{\sigma}_n(s,\overline{X}^n_s) d\hat{W}_s||_{L^2(\Omega)} \leq &C\sqrt{t}\sup_{t \in [0,T]}||\lambda(t,\overline{X}_t) - \lambda_n(t,\overline{X}_t^n)||_{L^2(\Omega)} \\
\leq & C \sqrt{t}\left(\P(\tau_n \leq T)^{1/4} + \sup_{t \in [0,T]}||\overline{X}_t - \overline{X}_t^n||_{L^2(\Omega)} \right).
\end{align*}
Hence,
\[
\sup_{t \in [0,T]} ||\overline{X}_t - \overline{X}_t^n||_{L^2(\Omega)} \leq C \sqrt{T}\left( \P(\tau_n \leq T)^{1/4} + \sup_{t \in [0,T]} ||\overline{X}_t - \overline{X}_t^n||_{L^2(\Omega)}\right)
\]
and therefore, for $T$ small enough, there exists $C > 0$ such that
\begin{equation} \label{eq: bound difference}
\sup_{t \in [0,T]} ||\overline{X}_t - \overline{X}_t^n||_{L^2(\Omega)} \leq C \P(\tau_n \leq T)^{1/4}.
\end{equation}
Thus,
\begin{align*}
||I_T^{LV}-I_T^{LV,n} ||_{L^2(\Omega)} \leq A(T) + C T \sup_{t \in [0,T]}||\overline{X}_t- \overline{X}_t^n||_{L^2(\Omega)} \leq CT \P(\tau_n \leq T)^{1/4}.
\end{align*}
Using the bound in Lemma \ref{lemma: bound stopping time} with $H = \frac{1}{2}$ and $p \geq 2$ large enough, we find that
\[
\lim_{T \to 0} \frac{||I_T^{LV}-I_T^{LV,n} ||_{L^2(\Omega)} }{T^4} = 0
\]
and therefore
\[
\lim_{T \to 0} \frac{|\E[(I_T^{LV})^2] - \E[(I_T^{LV,n})^2]|}{T^4} = 0,
\]
leading to the fact that if the constant $K(\alpha,\beta,\rho)$ defined in Theorem \ref{th: main}-(ii) is negative, then
\[
\lim_{T \to 0} \frac{\E[(I_T)^2]-\E[(I_T^{LV})^2]}{T^4}  < 0
\]
for $v$ satisfying (H'1) with $H = \frac{1}{2}$.
\subsection{Examples: The rough case}
In the previous section we have seen that if $H < \frac{1}{2}$ and the variance process satisfies (H'1), the Gatheral conjecture can not hold true if
\[
\rho^2  > \frac{2H+3}{4}.
\]
Consider $v$ satisfying the rough Heston model, that is,
\[
v_t = v_0 + \kappa\int_0^t (\theta-v_s)K(t-s) ds + \nu \int_0^t \sqrt{v_s} K(t-s) dW_s, \quad K(t) = \frac{t^{H-\frac{1}{2}}}{\Gamma(H+1/2)}.
\]
 Notice that, if $v_0 = \theta$ then $K^* = \E[I_T] = \theta T$, so we will consider $v_ 0 = \theta$ so the ATM strike can be computed analytically. We choose $H = 0.1$, so the Gatheral conjecture fails when
 \[
 |\rho| > \frac{\sqrt{2H+3}}{2} \approx 0.894427.
 \]
We select the parameters $\kappa = 0.3$, $\nu = 0.2$, $v_0 = \theta = 0.04$ and $T = 0.01$. We take $\rho = -0.925$, $\rho = 0.95$ and $\rho = -1$. We let
\[
\mathcal{C} := \E[(I_T-K^*)_+], \quad \mathcal{C}^{LV} = \E[(I_T^{LV}-K^*)_+].
\]
In order to simulate the local variance, we simulate an implied volatility surface under the rough Heston model in a fine grid. Then, we obtain several values $\lambda(t,x)$ using Dupire's formula. The quantity $\lambda(t,\overline{X}_t)$ is obtained by interpolation. The results are displayed in Table \ref{tab: rough Heston}.
\begin{table}[ht]
  \centering

  \begin{tabular}{ccccc}
    \toprule
    $T$ & $\rho$ & $\mathcal{C}$ & $\mathcal{C}^{LV}$ & $\mathcal{C}/\mathcal{C}^{LV}$ \\
    \midrule
    $0.01$ & $-0.925$ & $7.276422 \cdot 10^{-5}$ & $7.512216 \cdot 10^{-5}$ & $0.968612$ \\
    $0.01$ & $0.95$   & $7.276422 \cdot 10^{-5}$ & $7.612050 \cdot 10^{-5}$ & $0.955908$ \\
    $0.01$ & $-1$     & $7.276422 \cdot 10^{-5}$ & $8.127028 \cdot 10^{-5}$ & $0.895336$ \\
    \bottomrule
  \end{tabular}
    \caption{Study of the quotient $\mathcal{C}/\mathcal{C}^{LV}$ for different correlation values $\rho$.}
  \label{tab: rough Heston}
\end{table}
\subsection{Examples: The Markovian case}
From Theorem \ref{th: main}-(ii), in order to prove that a stochastic volatility model with $|\rho| = 1$ does not satisfy the Gatheral conjecture it is enough to check if the constant $K_0(\alpha,\beta,\rho) < 0$. We will compute $K_0(\alpha,\beta,\rho)$ for the Heston model and the full lognormal SABR model. We recall that
\[
K_0(\alpha,\beta,\rho) = \alpha'_0 \beta_0^2+ \frac{\alpha_0\beta_0^2}{2v_0} - \frac{\beta_0^4}{8v_0^2} - \frac{\rho \beta_0^3}{2\sigma_0}-\alpha_0 \beta_0 \beta_0' - \frac{\beta_0^3\beta_0^{''}}{2}
\]
\subsubsection{The Heston model}
Let $|\rho| = 1$, that is, $\rho \in \{-1,1\}$ and let $v$ satisfy the SDE
\[
dv_t = \kappa(\theta-v_t) dt + \nu\sqrt{v_t}dW_t.
\]
Then, $v$ is of the form $dv_t = \alpha(v_t)dt + \beta(v_t)dW_t$ with $\alpha(v) = \kappa(\theta-v)$ and $\beta(v) = \nu\sqrt{v}$. In order to compute $K_0(\alpha,\beta,\rho)$, observe that
\begin{itemize}
    \item $\alpha_0 = \kappa (\theta-v_0)$.
    \item $\alpha'_0 = -\kappa$.
    \item $\beta_0 = \nu \sigma_0$ (recall $\sigma_0 = \sqrt{v_0}$),
    \item $\beta'_0 = \frac{\nu}{2\sigma_0}$.
    \item $\beta_0'' = \frac{-\nu}{4\sigma_0^3}$.
\end{itemize}
Thus,
\[
\alpha'_0 \beta_0^2+ \frac{\alpha_0\beta_0^2}{2v_0} - \frac{\beta_0^4}{8v_0^2} - \frac{ \rho\beta_0^3}{2\sigma_0}-\alpha_0 \beta_0 \beta_0' - \frac{\beta_0^3\beta_0^{''}}{2} = \nu^2 v_0\left(\frac{\rho\nu}{2}-\kappa\right).
\]
Hence, we find that if $\rho = -1$, then
\[
K_0(\alpha,\beta,-1) = \nu^2 v_0\left(\frac{-\nu}{2}-\kappa\right) < 0
\]
for all combination of parameters. However, if $\rho = 1$ then
\[
K_0(\alpha,\beta,1) = \nu^2 v_0\left(\frac{\nu}{2}-\kappa\right) < 0
\]
if $\nu < 2\kappa$.
\subsubsection{The SABR model}
Consider now the SABR model, which the volatility process is defined by
\[
d\sigma_t = \alpha \sigma_t dW_t.
\]
In this case we need to translate the SDE to the variance with the Itô formula.
\[
dv_t = d(\sigma_t^2) = 2\sigma_t d\sigma_t + d\langle \sigma\rangle_t = \alpha^2 v_t dt + 2\alpha v_t dW_t.
\]
We have
\begin{itemize}
    \item $\alpha_0 = \alpha^2 v_0$.
    \item $\alpha'_0 = \alpha^2$.
    \item $\beta_0 = 2\alpha v_0$.
    \item $\beta'_0 = 2\alpha$.
    \item $\beta''_0 = 0$.
\end{itemize}
Thus, the constant is
\[
\alpha'_0 \beta_0^2+ \frac{\alpha_0\beta_0^2}{2v_0} - \frac{\beta_0^4}{8v_0^2} - \frac{ \beta_0^3}{2\sigma_0}-\alpha_0 \beta_0 \beta_0' - \frac{\beta_0^3\beta_0^{''}}{2} = 4\rho\alpha^3 \sigma_0^5 = 4\rho\alpha^3 v_0^{5/2}.
\]
In this case, the refutation of Gatheral's conjecture depends only on the sign of $\rho$. That is, if $\rho = -1$ the conjecture does not hold and, if $\rho = 1$, the short-time behavior of the difference between quadratic payoffs is consistent with Gatheral's conjecture. For a numerical experiment, consider the SABR model with parameters $\alpha = 1.20$, $\rho = -1$ and $T = 0.01$. In order to simulate the local variance, we use the same approach as with the rough Heston case. That is, we simulate an implied volatility surface under the SABR model using Hagan's formula and we obtain $\lambda(t,x)$ in a grid using Dupire's formula. The results are displayed in Table \ref{tab: SABR}.
\begin{table}[ht]
  \centering
  \begin{tabular}{cccc}
    \toprule
    $\E[I_T^2]$ & $\E[(I_T^{LV})^2]$
    & $\frac{\E[I_T^2]-\E[(I_T^{LV})^2]}{T^4}$
    & MC $95\%$ CI\\
    \midrule
    $1.653199\times10^{-7}$
    & $1.653233\times10^{-7}$
    & $-0.000344$
    & $[-0.000457,-0.000230]$ \\
    \bottomrule
  \end{tabular}
  \caption{Study of the difference
    $(\E[I_T^2]-\E[(I_T^{LV})^2])/T^4$ for the SABR model.}
  \label{tab: SABR}
\end{table}

\section{Conclusions}
In this paper, we have provided sufficient conditions for a general stochastic volatility model to refute Gatheral's conjecture based on the short-time properties of integrated variance derivatives and Wiener chaos expansions of both the stochastic and the local variance. The main conclusions of our main result, that is Theorem \ref{th: main}, state that for a general class of rough volatility models, Gatheral's conjecture can't hold if the correlation is higher than a threshold that depends on the Hurst index. For Markovian variances which can be expressed as SDEs, we show that Gatheral's conjecture can't hold true if a certain relationship between the coefficients of the SDE is met.

\section*{Declaration of the use of generative AI}
During the preparation of the manuscript, the authors used OpenAI's ChatGPT Astra 6 to verify the work once a first version of the paper was ready. ChatGPT Astra 6 contributed to filling some gaps in the localization argument. The idea of taking the quotient of the implied volatilities and using Wiener chaos expansions for the Markovian case was developed by the authors. The authors take full responsibility for the AI-assisted content in the paper. ChatGPT Astra 6 also detected several typographical errors.
\section*{Acknowledgments}
Author Òscar Burés is supported by program AGAUR-FI ajuts (2025 FI-1 00580) from the Department of Research and Universities of the Government of Catalonia and the co-funding of the European Social Fund Plus (ESF+).
\appendix

\section{The local variance model} \label{sec: local variance model}
Let $X$ be the log-price of the stochastic volatility model
\[
dX_t = -\frac{1}{2}v_t dt + \sqrt{v_t}(\rho dW_t + \sqrt{1-\rho^2}dB_t
\]
and let
\begin{equation} \label{eq: local variance appendix}
d\overline{X}_t = -\frac{1}{2}\lambda(t,\overline{X}_t)dt + \sqrt{\lambda(t,\overline{X}_t)} d\hat{W}_t
\end{equation}
denote the log-price of the associated local variance of the model. In this section we want to detail the existence, uniqueness and Malliavin differentiability of $\overline{X}$.
\begin{prop}
    There exists a unique process $\overline{X}$ that solves the SDE given by Equation \eqref{eq: local variance appendix}. Moreover, $\overline{X}_t \in \D^{1,p}$ for all $p \geq 2$ and $t \in [0,T]$.
\end{prop}
\begin{proof}
    The existence and uniqueness of solutions is a consequence of \cite{lan2014new}. The Malliavin differentiability is discussed in \cite{AlosGLPravosud}.
\end{proof}

\section{Asymptotic expansions in the $H = \frac{1}{2}$ case}
In this section we let $v$ satisfy Equation \eqref{eq: variance model} with $H = \frac{1}{2}$. This section is devoted to state and proof all the asymptotic properties of $v$ and related processes needed in order to carry out the analysis in Section \ref{sec: Markovian case}. Therefore, in this section we will assume $v$ is of the form
\[
dv_t = \alpha(v_t)dt + \beta(v_t) dW_t
\]
with $\alpha, \beta \in \mathcal{C}_b^{\infty}(\R)$. In order to streamline the notation, for any $n \geq 0$ we will denote $\alpha^{(n)}_0$ the $n$-th derivative of $\alpha$ evaluated at $v_0$ (same for $\beta$). In the case of the first and second derivative we use the standard notations $\alpha'_0$ and $\alpha''_0$.
\subsection{Asymptotic results for $v_t$}
Write
\[
(\mathcal{L}f)(v) := \alpha(v)f'(v) + \frac{1}{2}\beta(v)^2 f''(v), \quad \theta := (\mathcal{L}\sqrt{\cdot})(v_0)
\]
\begin{prop} \label{prop: v properties}
    The following properties hold:
    \begin{itemize}
        \item[($v$-I)] $||v_t- v_0||_{L^p(\Omega)} \leq C_p t^{1/2}$.
        \item[($v$-II)] $\beta(v_t) = \beta_0 + \beta_0'\beta_0 W_t + O_{L^2(\Omega)}(t)$.
        \item[($v$-III)] $v_t= \E[v_t] + \beta_0W_t + \frac{\beta_0\beta_0'}{2}(W_t^2-t) + O_{L^2(\Omega)}(t^{3/2})$, with $\E[v_t] = v_0 + \alpha_0t + O(t^{3/2})$.
        \item[($v$-IV)] $||v_t^{\geq 2}||_{L^2(\Omega)} \leq Ct$.
        \item[($v$-V)] $\sqrt{v_t} = \sigma_0 + \frac{\beta_0}{2\sigma_0}W_s + O_{L^2(\Omega)}(s)$ and $\E[\sqrt{v_t}] =\sigma_0 + \theta t  + O(t^{3/2})$
    \end{itemize}
\end{prop}
\begin{proof}
    \leavevmode
    \begin{itemize}
        \item ($v$-I) Since $\alpha$ is bounded,
        \[
        ||v_t- v_0||_{L^p(\Omega)} \leq C_p t + ||\int_0^t \beta(v_s) dW_s||_{L^p(\Omega)}.
        \]
        The result follows immediately from the Burkholder-Davis-Gundy inequality and the fact that $\beta$ is bounded.
        \item[($v$-II)] The Itô formula applied to $\beta(v_t)$ gives
        \[
        \beta(v_t) = \beta_0 + \int_0^t (\mathcal{L}\beta)(v_s) ds +  \int_0^t \beta'(v_s)\beta(v_s) dW_s
        \]
        Since $\alpha$ and $\beta$ are smooth, we have
        \[
        \int_0^t (\mathcal{L}\beta)(v_s) ds = O(t).
        \]
        For the diffusion term,
        \[
        \int_0^t \beta'(v_s)\beta(v_s) dW_s = \beta_0 \beta'_0 W_t + \int_0^t [\beta(v_s)\beta'(v_s) - \beta_0 \beta'_0 ] dW_s
        \]
        Since the map $\beta\beta'$ is smooth and bounded, 
        \[
        |\beta(v_s)\beta'(v_s) - \beta_0 \beta'_0| \leq C|v_s-v_0|.
        \]
        Using the Itô isometry and the property ($v$-I) with $p = 2$ shows that
        \[
        ||\int_0^t [\beta(v_s)\beta'(v_s) - \beta_0 \beta'_0 ] dW_s||_{L^2(\Omega)} \leq Ct.
        \]
        \item[($v$-III)] We write
        \[
        v_t - \E[v_t] = \int_0^t (\alpha(v_s)-\E[\alpha(v_s)])ds + \int_0^t \beta(v_s) dW_s.
        \]
        On the one hand
        \[
        ||\alpha(v_s) - \E[\alpha(v_s)]||_{L^2(\Omega)} \leq ||\alpha(v_s) - \alpha_0 ||_{L^2(\Omega)} \leq C||v_s-v_0||_{L^2(\Omega)} \leq Cs^{1/2}
        \]
        so
        \[
        ||\int_0^t (\alpha(v_s)-\E[\alpha(v_s)])ds||_{L^2(\Omega)} \leq Ct^{3/2}.
        \]
        On the other hand, we use ($v$-II) to write
        \[
        \int_0^t \beta(v_s) dW_s = \beta_0 W_s + \frac{\beta_0\beta_0'}{2}(W_t^2 - t) + \int_0^tR_s dW_s
        \]
        with $||R_s||_{L^2(\Omega)} \leq Cs$. The Itô isometry gives us
        \[
        || \int_0^tR_s dW_s||_{L^2(\Omega)} \leq Ct^{3/2}.
        \]
        For $\E[v_t]$ we use the mean value theorem and ($v$-I) to write
        \[
        |\E[v_t] - v_0 - \alpha_0 t|  \leq \int_0^t |\E[\alpha(v_s) - \alpha_0]| ds \leq C\int_0^t ||v_s-v_0||_{L^2(\Omega)} ds \leq Ct^{3/2}.
        \]
        \item[($v$-IV)] We use ($v$-III) to write
        \[
        v^{\geq 2}_t = \frac{\beta_0 \beta_0'}{2}(W_t^2 -t)^{\geq 2} + R_t^{\geq 2}.
        \]
        On the one hand,
        \[
        ||(W_t^2 -t)^{\geq 2}||_{L^2(\Omega)} \leq ||W_t^2 -t ||_{L^2(\Omega)} \leq Ct.
        \]
        On the other hand,
        \[
        ||R_t^{\geq2}||_{L^2(\Omega)} \leq ||R_t||_{L^2(\Omega)} \leq Ct^{3/2}.
        \]
        Thus,
        \[
        ||v_t^{\geq 2}||_{L^2(\Omega)} \leq Ct.
        \]
        \item[($v$-V)] The proof follows the same lines as in ($v$-II) with the function $\sqrt{v_t}$ instead of $\beta(v_t)$.
    \end{itemize}
\end{proof}

\subsection{Asymptotics results for $D_sv_t$}

Recall that for any $0 \leq s < t \leq T$ we have
\[
D_sv_t = \beta(v_s) + \int_s^t \alpha'(v_u) D_sv_u du + \int_s^t \beta'(v_u)D_sv_u dW_u.
\]

\begin{prop} \label{prop: Dv properties}
    The following properties hold:
    \begin{itemize}
        \item[($Dv$-I)] We have
        \[
        \sup_{r,s \leq t} \left(||D_sv_t||_{L^p(\Omega)} + ||D_rD_sv_t||_{L^p(\Omega)} \right) \leq C_p.
        \]
        \item[($Dv$-II)] We have 
        \[
        ||D_sv_t - \beta_0||_{L^2(\Omega)} \leq C t^{1/2}
        \]
        and, as a consequence, $ ||D_sv_t - \E[D_sv_t]||_{L^2(\Omega)} \leq Ct^{1/2}$.
        \item[($Dv$-III)] We have
        \[
        \E[D_sv_t] = \beta_0 + \alpha'_0 \beta_0(t-s) + (\mathcal{L}\beta)(v_0)s + O(t^{3/2}).
        \]
        \item[($Dv$-IV)] We have
        \[
        ||D_s\sqrt{v_t} - \frac{\beta_0}{2\sigma_0}||_{L^2(\Omega)} \leq Ct^{1/2}.
        \]
    \end{itemize}
\end{prop}

\begin{proof}
    \leavevmode
    \begin{itemize}
        \item[($Dv$-I)] The proof follows a standard Gronwall argument. See for instance \cite[Theorems 2.2.1 and 2.2.2]{nualart2006}.
        \item[($Dv$-II)] We have
        \[
        D_sv_t - \beta_0 = (\beta(v_s)-\beta_0) +\int_s^t \alpha'(v_u) D_sv_u du + \int_s^t \beta'(v_u)D_sv_u dW_u.
        \]
        The mean value theorem and ($v$-I) shows that $||\beta(v_s) - \beta_0||_{L^p(\Omega)} \leq Ct^{1/2}$. Property ($Dv$-I) together with Holder's inequality and Burkholder-Davis-Gundy inequality yield respectively
        \[
        ||\int_s^t \alpha'(v_u)D_sv_udu||_{L^p(\Omega)} \leq Ct, \quad ||\int_s^t \beta'(v_u)D_sv_u dW_u||_{L^p(\Omega)} \leq Ct^{1/2}.
        \]
        Now, since $||D_sv_t - \E[D_sv_t]||_{L^p(\Omega)} \leq ||D_sv_t - \beta_0||_{L^p(\Omega)}$ we conclude the desired estimate.
        \item[($Dv$-III)] Taking expectations we have
        \[
        \E[D_sv_t] = \E[\beta(v_s)] + \int_s^t \E[\alpha'(v_u) D_sv_u] du.
        \]
        The Itô formula on $\beta(v_s)$ in the same way as in the proof of ($v$-II) yields
        \[
        \E[\beta(v_s)] = \beta_0 + \int_0^s \E[ (\mathcal{L}\beta)(v_r) ]dr.
        \]
        Since the map $\mathcal{L}\beta$ is smooth we have that
        \[
        \E[|(\mathcal{L}\beta)(v_r) - (\mathcal{L}\beta)_0|] \leq Cr^{1/2}.
        \]
        Integrating,
        \begin{equation} \label{eq: Dv-III contribution beta}
        \E[\beta(v_s)] = \beta_0 + (\mathcal{L}\beta)_0 s + O(s^{3/2}).
        \end{equation}
        Finally,
        \[
        \E[\alpha'(v_u)D_sv_u]-\alpha'_0 \beta_0 = \E[D_sv_u(\alpha'(v_u) - \alpha'_0)] + \E[\alpha'(v_u)(D_rv_u - \beta_0)]
        \]
        Using the smoothness on $\alpha$ and the properties ($Dv$-I) and ($Dv$-II) we see that
        \begin{equation} \label{eq: Dv-III contribution derivative}
        \int_s^t \E[\alpha'(v_u) D_sv_u] du = \alpha'_0\beta_0 (t-s) + O(t^{3/2}).
        \end{equation}
        Adding the contributions \eqref{eq: Dv-III contribution beta} and \eqref{eq: Dv-III contribution derivative} we obtain the result.
        \item[($Dv$-IV)] The chain rule for Malliavin calculus yields
        \[
        D_s\sqrt{v_t} = \frac{D_sv_t}{2\sqrt{v_t}}.
        \]
        Hence,
        \[
        D_s\sqrt{v_t} - \frac{\beta_0}{2\sigma_0} = \frac{D_sv_t - \beta_0}{2\sqrt{v_t}} + \frac{\beta_0}{2} \left( \frac{1}{\sqrt{v_t}} - \frac{1}{\sigma_0} \right) =: T_1  + T_2
        \]
        For $T_2$, since $v$ is bounded from below and the fact function $f(x) = x^{-1}$ is Lipschitz continuous on any interval $[M, \infty)$ yields that
        \[
        ||T_2||_{L^2(\Omega)} \leq C||v_t - v_0 ||_{L^2(\Omega)} \leq Ct^{1/2{}}
        \]
        because of ($v$-I). For $T_1$ we also use that $v$ is bounded from below to conclude that
        \[
        ||T_1 ||_{L^2(\Omega)} \leq C||D_sv_t - \beta_0||_{L^2(\Omega)} \leq Ct^{1/2}
        \]
        due to ($Dv$-II). Joining the estimates of $T_1$ and $T_2$ we obtain the result.
    \end{itemize}
\end{proof}

\subsection{Asymptotic results for $X_t = \log S_t$}

\begin{prop} \label{prop: X properties}
    Let $X_t^c = X_t - \E[X_t]$. The following properties hold:
    \begin{itemize}
        \item[($X$-I)] $||X_t^c||_{L^p(\Omega)} \leq C_pt^{1/2}$ for all $p \geq 2$.
        \item[($X$-II)] $X_t^c = \rho \sigma_0 W_t + \frac{\rho \beta_0}{4\sigma_0}(W_t^2 -t) +O_{L^2(\Omega)}(t^{3/2})$. Moreover,
        \[
        ||X_t^c - \rho\sigma_0 W_t||_{L^p(\Omega)} \leq Ct
        \]
        for all $p \geq 2$.
        \item[($X$-III)] $||(X_t^c)^{\geq 2}||_{L^2(\Omega)} \leq Ct$.
        \item[($X$-IV)] $(X_t^c)^2 - \E[(X_t^c)^2] = v_0 (W_t^2 -t) + O_{L^2(\Omega)}(t^{3/2}).$
    \end{itemize}
\end{prop}
\begin{proof}
    \leavevmode
    \begin{itemize}
        \item [($X$-I)] Write
        \[
        X_t^c = -\frac{1}{2}\int_0^t (v_s - \E[v_s])ds + \rho \int_0^t \sqrt{v_s} dW_s.
        \]
        The result follows from ($v$-III) and the fact that $v_{min} \leq v_t \leq v_{max}$.

        \item[($X$-II)] On the one hand, we use ($v$-III) to estimate
        \[
        ||\int_0^t (v_s - \E[v_s])ds||_{L^2(\Omega)} \leq \int_0^t ||v_s-\E[v_s]||_{L^2(\Omega)}ds \leq Ct^{3/2}.
        \]
        For the diffusion term, we use ($v$-V) to write
        \[
       \rho \int_0^t \sqrt{v_s} dW_s = \rho\sigma_0 W_t + \frac{\rho\beta_0}{4\sigma_0}(W_t^2 - t) + \int_0^t R_s dW_s
        \]
        with $||R_t||_{L^2(\Omega)} \leq Ct$. The result follows from the Itô isometry.
        \item[($X$-III)] From ($X$-II) we have
        \[
        ||(X_t^c)^{\geq 2}||_{L^2(\Omega)} \leq C||W_t^2 - t||_{L^2(\Omega)} \leq Ct.
        \]
        \item[($X$-IV)] Notice that 
        \[
        v_0 (W_t^2 -t) = v_0W_t^2 - \E[v_0W_t^2],
        \]
        so
        \[
        (X_t^c)^2 - \E[(X_t^c)^2] - v_0(W_t^2 -t) = (X_t^c - v_0 W_t^2) - \E[(X_t^c)^2 - v_0 W_t^2].
        \]
        Thus,
        \[
        ||(X_t^c)^2 - \E[(X_t^c)^2] - v_0(W_t^2 -t)||_{L^2(\Omega)} \leq ||(X_t^c)^2 - v_0W_t^2 ||_{L^2(\Omega)}.
        \]
        Since $(X_t^c)^2- v_0W_t^2 = (X_t^c - \rho \sigma_0W_t)(X_t^c + \rho\sigma_0W_t)$ we have, from Hölder's inequality,
        \[
         ||(X_t^c)^2 - v_0W_t^2 ||_{L^2(\Omega)} \leq ||X_t^c - \rho \sigma_0 W_t||_{L^4(\Omega)} ||X_t^c+ \rho\sigma_0 W_t||_{L^4(\Omega)}.
        \]
        On the one hand, ($X$-I) yields $||X_t^c+ \rho\sigma_0 W_t||_{L^4(\Omega)} \leq Ct^{1/2}$. On the other hand, ($X$-II) shows that $||X_t^c - \rho \sigma_0 W_t||_{L^4(\Omega)} \leq Ct$. The product of the estimates yields
        \[
          ||(X_t^c)^2 - \E[(X_t^c)^2] - v_0(W_t^2 -t)||_{L^2(\Omega)}\leq Ct^{3/2}.
        \]
    \end{itemize}
\end{proof}
\subsection{Asymptotic results for $D_sX_t$} 
Recall that
\[
D_sX_t = \rho\sqrt{v_s} - \frac{1}{2}\int_s^t D_sv_u du + \rho\int_s^t D_s\sqrt{v_u} dW_u.
\]
\begin{prop} \label{prop: DX properties}
    The following properties hold:
    \begin{itemize}
        \item [($DX$-I)] For every $p \geq 2$, there exists $C_p > 0$ such that $\sup_{s \leq t} ||D_sX_t||_{L^p(\Omega)} \leq C$.
        \item[($DX$-II)] We have
        \[
        \E[D_sX_t] = \rho \sigma_0 + \rho \theta s - \frac{\beta_0}{2}(t-s) + O(t^{3/2}).
        \]
        \item[($DX$-III)] We have
        \[
        D_sX_t - \E[D_sX_t] = \frac{\rho \beta_0}{2\sigma_0}W_t + R_{s,t}, \quad ||R_{s,t} ||_{L^2(\Omega)} \leq Ct.
        \]
    \end{itemize}
\end{prop}

\begin{proof}
    \leavevmode
    \begin{itemize}
        \item [($DX$-I)] From ($v$-I) and ($Dv$-I) we conclude that
        \[
        \sup_{r \leq t} \left( ||\sqrt{v_r}||_{L^p(\Omega)} + ||\int_s^t D_sv_u du ||_{L^p(\Omega)}\right) \leq C_p.
        \]
        For the stochastic integral we obtain the same bound using Burkholder-Davis-Gundy together with ($Dv$-IV).
        \item[($DX$-II)] We have
        \[
        \E[D_sX_t] = \rho \E[\sqrt{v_s}] - \frac{1}{2}\int_s^t\E[D_sv_u] du.
        \]
        Plugging ($v$-V) and ($Dv$-III) into the previous equation yields
        \[
        \E[D_sX_t] = \rho \sigma_0 + \rho \theta s -\frac{1}{2} \int_s^t (\beta_0 + \alpha'\beta_0 (u-s) + (\mathcal{L}\beta)_0 s + O(u^{3/2})) du + O(s^{3/2}).
        \]
        Integrating,
        \[
        \E[D_sX_t] = \rho \sigma_0 + \rho \theta s - \frac{1}{2}\beta_0(t-s) + O(t^{3/2}).
        \]
        \item[($DX$-III)] We write
        \[
        D_sX_t - \E[D_sX_t]= T_1 + T_2 + T_3,
        \]
        where
        \[
        T_1 = \rho(\sqrt{v_s}-\E[\sqrt{v_s})], \quad T_2 = -\frac{1}{2}\int_s^t (D_sv_u - \E[D_sv_u]) du,
        \]
        and
        \[
        T_3 = \rho \int_s^t D_s\sqrt{v_u}dW_u.
        \]
        For $T_1$, we want to show that
        \[
        T_1 = \frac{\rho \beta_0}{2\sigma_0}W_s + O_{L^2(\Omega)}(t).
        \]
        To do so, we use ($v$-V) to estimate
        \[
        ||T_1 -\frac{\rho \beta_0}{2\sigma_0}W_s ||_{L^2(\Omega)} = ||\rho (\sqrt{v_s}- \E[\sqrt{v_s}]) - \frac{\rho \beta_0}{2\sigma_0}W_s ||_{L^2(\Omega)} \leq Cs  \leq Ct.
        \]
        For $T_2$ we use ($Dv$-II) to write
        \[
        ||T_2||_{L^2(\Omega)} \leq C \int_s^t ||D_sv_u - \E[D_sv_u] ||_{L^2(\Omega)} du \leq Ct^{3/2}.
        \]
        For $T_3$ we rewrite
        \[
        \rho\int_s^t D_s\sqrt{v_u} dW_u = \rho\int_s^t\left(D_s\sqrt{v_u} - \frac{\beta_0}{2\sigma_0} \right)d W_u  + \frac{\rho\beta_0}{2\sigma_0}(W_t-W_s).
        \]
        Using Burkholder-Davis-Gundy inequality and ($Dv$-IV) we have that
        \[
           \rho\int_s^t D_s\sqrt{v_u} dW_u  = \frac{\rho\beta_0}{2\sigma_0}(W_t-W_s) + O_{L^2(\Omega)}(t).
        \]
        Adding the contributions from $T_1$, $T_2$ and $T_3$ we find that
        \begin{align*}
           D_sX_t - \E[D_sX_t] = &\frac{\rho\beta_0}{2\sigma_0}W_s + \frac{\rho\beta_0}{2\sigma_0}(W_t-W_s) + O_{L^2(\Omega)}(t) \\
           = & \frac{\rho\beta_0}{2\sigma_0}W_t + O_{L^2(\Omega)}(t),
        \end{align*}
        which concludes the result.
    \end{itemize}
\end{proof}

\bibliographystyle{apalike}
\bibliography{references.bib}

@book {nualart2006,
    AUTHOR = {Nualart, David},
     TITLE = {The {M}alliavin calculus and related topics},
    SERIES = {Probability and its Applications (New York)},
   EDITION = {Second},
 PUBLISHER = {Springer-Verlag, Berlin},
      YEAR = {2006},
     PAGES = {xiv+382},
      ISBN = {978-3-540-28328-7; 3-540-28328-5},
   MRCLASS = {60-02 (60H07 60H30)},
  MRNUMBER = {2200233},
MRREVIEWER = {Daniel\ Ocone},
}

@article {AlosGLPravosud,
    AUTHOR = {Al\`os, Elisa and Garc\'{\i}a-Lorite, David and Pravosud,
              Makar},
     TITLE = {On the skew and curvature of the implied and local
              volatilities},
   JOURNAL = {Appl. Math. Finance},
  FJOURNAL = {Applied Mathematical Finance},
    VOLUME = {30},
      YEAR = {2023},
    NUMBER = {1},
     PAGES = {47--67},
      ISSN = {1350-486X,1466-4313},
   MRCLASS = {91B70 (60H07)},
  MRNUMBER = {4653896},
MRREVIEWER = {Indranil\ SenGupta},
       DOI = {10.1080/1350486X.2023.2261459},
       URL = {https://doi.org/10.1080/1350486X.2023.2261459},
}

@article{alosGLmuguruza,
  title={On smile properties of volatility derivatives: Understanding the VIX skew},
  author={Alos, Elisa and Garc{\'\i}a-Lorite, David and Gonzalez, Aitor Muguruza},
  journal={SIAM Journal on Financial Mathematics},
  volume={13},
  number={1},
  pages={32--69},
  year={2022},
  publisher={SIAM}
}

@article{bures2026skew,
  title={On the skew-curvature power law relationship in a stochastic volatility model with infinite variation {L}{\'e}vy jumps},
  author={Bur{\'e}s, {\`O}scar and Vives, Josep},
  journal={arXiv preprint arXiv:2610.23488},
  year={2026}
}

@article{lucic2026gatheral,
  title={Gatheral's Conjecture Revisited},
  author={Lucic, Vladimir},
  journal={arXiv preprint arXiv:2609.05047},
  year={2026}
}

@article{chen2021spatial,
  title={Spatial ergodicity for SPDEs via Poincar{\'e}-type inequalities},
  author={Chen, Le and Khoshnevisan, Davar and Nualart, David and Pu, Fei},
  journal={Electronic Journal of Probability},
  volume={26},
  pages={1--37},
  year={2021},
  publisher={The Institute of Mathematical Statistics and the Bernoulli Society}
}

@article{alos2007short,
  title={On the short-time behavior of the implied volatility for jump-diffusion models with stochastic volatility},
  author={Alos, Elisa and Le{\'o}n, Jorge A and Vives, Josep},
  journal={Finance and stochastics},
  volume={11},
  number={4},
  pages={571--589},
  year={2007},
  publisher={Springer}
}

@article{lan2014new,
  title={New sufficient conditions of existence, moment estimations and non confluence for SDEs with non-Lipschitzian coefficients},
  author={Lan, Guangqiang and Wu, Jiang-Lun},
  journal={Stochastic Processes and their Applications},
  volume={124},
  number={12},
  pages={4030--4049},
  year={2014},
  publisher={Elsevier}
}

@article{ou1957boundedness,
  title   = {The boundedness of solutions of linear differential equations {$y''+A(t)y=0$}},
  author  = {Ou-Iang, Liang},
  journal = {Shuxue Jinzhan (Advances in Mathematics)},
  volume  = {3},
  pages   = {409--415},
  year    = {1957}
}

@article{cheung2006discrete,
  title={Discrete non-linear inequalities and applications to boundary value problems},
  author={Cheung, Wing-Sum and Ren, Jingli},
  journal={Journal of Mathematical Analysis and Applications},
  volume={319},
  number={2},
  pages={708--724},
  year={2006},
  publisher={Elsevier}
}

@article{gatheral2005valuation,
  title={Valuation of volatility derivatives},
  author={Gatheral, Jim},
  journal={Global Derivatives and Risk Management},
  volume={24},
  year={2005}
}

@article{brockhaus2000volatility,
  title={Volatility swaps made simple},
  author={Brockhaus, Oliver and Long, Douglas},
  journal={RISK-LONDON-RISK MAGAZINE LIMITED-},
  volume={13},
  number={1},
  pages={92--95},
  year={2000},
  publisher={RISK MAGAZINE LIMITED}
}

@article{beiglbock2011minimum,
  title={Is the minimum value of an option on variance generated by local volatility?},
  author={Beiglb{\"o}ck, Mathias and Friz, Peter and Sturm, Stephan},
  journal={SIAM Journal on Financial Mathematics},
  volume={2},
  number={1},
  pages={213--220},
  year={2011},
  publisher={SIAM}
}

@article{BrunickShreve2013,
  author  = {Brunick, Gerard and Shreve, Steven},
  title   = {Mimicking an {I}t{\^o} process by a solution of a stochastic differential equation},
  journal = {The Annals of Applied Probability},
  volume  = {23}, number = {4}, pages = {1584--1628}, year = {2013}}

@article{alfonsi2025nonnegativity,
  title={Nonnegativity preserving convolution kernels. Application to Stochastic Volterra Equations in closed convex domains and their approximation},
  author={Alfonsi, Aur{\'e}lien},
  journal={Stochastic Processes and their Applications},
  volume={181},
  pages={104535},
  year={2025},
  publisher={Elsevier}
}
\end{document}